\documentclass[twoside,leqno]{article}

\usepackage[T1]{fontenc}
\usepackage[utf8]{inputenc}
\usepackage[english]{babel}
\usepackage{amssymb,amsmath,amsfonts,amsthm}

\newtheorem{theorem}{Theorem}
\newtheorem{definition}{Definition}
\newtheorem{lemma}{Lemma}
\newtheorem{corollary}{Corollary}

\usepackage{mathtools} % Bonus

\usepackage{amsfonts}

\usepackage{enumitem,fullpage}
\usepackage{amsopn}

\usepackage{mathcomp,tabularx,subfig}
\usepackage{multirow}
\usepackage{graphicx}
\usepackage{textcomp}
\usepackage{xcolor,cite}
\usepackage{comment}
\usepackage{xspace,tikz,makecell}
\usetikzlibrary{positioning}
\usetikzlibrary{calc}
\usepackage[algo2e,ruled,procnumbered,linesnumbered, noend]{algorithm2e}
\usepackage{booktabs} 
\usepackage{thmtools}
\usepackage{thm-restate,hyperref}
\hypersetup{
colorlinks=true
}

\def\bigO{\ensuremath{\mathcal{O}}\xspace}

\newcommand{\blocco}[1]{\par\noindent\textbf{#1.}}
\SetKwComment{tcp}{$\triangleright$ }{}%

\SetCommentSty{mycommfont}

\newcommand{\expsinglesource}{\textsc{\textmd{pruned-\ssksispshort}}\xspace}
\newcommand{\polysinglesource}{\textsc{\textmd{bounded-\ssksispshort}}\xspace}
\newcommand{\nout}{\ensuremath{N^+}\xspace}
\newcommand{\nin}{\ensuremath{N^-}\xspace}

\newcommand{\testpruning}{\ensuremath{\textsc{pruning}}\xspace}

\newcommand{\exhaustive}{\textsc{exh-\ssksispshort}\xspace}

\newcommand{\ssksisp}{Single-Source Top-$k$ Simple Shortest Paths\xspace}
\newcommand{\spksisp}{Single-Pair Top-$k$ Simple Shortest Paths\xspace}
\newcommand{\spkalgo}{\textsc{\textmd{\spksispshort-algo}}\xspace}

\newcommand{\ssksispshort}{\textsc{\textmd{S-kssp}}\xspace}
\newcommand{\spksispshort}{\textsc{\textmd{P-kssp}}\xspace}
\newcommand{\solss}{\ensuremath{{S}}\xspace} %
\newcommand{\solssvertex}[1]{\ensuremath{{S}_{r #1}}\xspace} %
\newcommand{\soluzionepair}[2]{\ensuremath{{S}_{#1#2}}\xspace}
\newcommand{\collezionesol}[1]{\ensuremath{\Gamma_{r #1}}\xspace} %

\newcommand{\wg}[1]{\ensuremath{\omega(#1)}\xspace}

\newcommand{\ssyen}{\textsc{ss-yen}\xspace}
\newcommand{\sspnc}{\textsc{ss-pnc}\xspace}
\newcommand{\pnc}{\textsc{pnc}\xspace}
\newcommand{\shortpolyyen}{\textsc{bnd-yen}\xspace}
\newcommand{\shortpolypnc}{\textsc{bnd-pnc}\xspace}

\newcommand{\shortspeedup}{\textsc{sp-u}\xspace}

\newcommand{\bestk}[1]{\ensuremath{\textsc{T}_{#1}}\xspace}
\newcommand{\checked}{\ensuremath{\textsc{seen}}\xspace}
\newcommand{\supersat}{\ensuremath{\textsc{s-sat}}\xspace}

\newcommand{\true}[2][black,fill=black]{\tikz[baseline=-0.5ex]\draw[#1,radius=#2] (0,0) circle ;}
\newcommand{\false}[2][black,fill=white]{\tikz[baseline=-0.5ex]\draw[#1,radius=#2] (0,0) circle ;}

\newcommand{\maxdeg}{\ensuremath{d_{\textsc{max}}}\xspace}

\newcommand{\avgdeg}{\ensuremath{d_{\textsc{avg}}}\xspace}

\newcommand{\oregon}{\textsc{Oregon-AS}\xspace}
\newcommand{\oregonss}{\textsc{ore}\xspace}
\newcommand{\caida}{\textsc{Caida}\xspace}
\newcommand{\caidass}{\textsc{cai}\xspace}
\newcommand{\bright}{\textsc{Brightkite}\xspace}
\newcommand{\brightss}{\textsc{bri}\xspace}
\newcommand{\citat}{\textsc{Ca-GrQc}\xspace}
\newcommand{\citatss}{\textsc{cag}\xspace}

\newcommand{\peer}{\textsc{P2p-Gnutella}\xspace}
\newcommand{\peerss}{\textsc{gnu}\xspace}

\newcommand{\direrdos}{\textsc{Erdős-Rényi}\xspace}
\newcommand{\direrdosss}{\textsc{erd}\xspace}

\newcommand{\linux}{\textsc{Linux}\xspace}
\newcommand{\linuxss}{\textsc{lin}\xspace}
\newcommand{\epi}{\textsc{Epinions}\xspace}
\newcommand{\episs}{\textsc{epi}\xspace}
\newcommand{\lux}{\textsc{Luxembourg}\xspace}
\newcommand{\luxss}{\textsc{lux}\xspace}

\newcommand{\slashd}{\textsc{SlashDot}\xspace}
\newcommand{\slashdss}{\textsc{sld}\xspace}

\newcommand{\moc}{\textsc{Mocnik}\xspace}
\newcommand{\mocss}{\textsc{moc}\xspace}

\newcommand{\scf}{\textsc{Scale-Free}\xspace}
\newcommand{\scfss}{\textsc{scf}\xspace}
\newcommand{\barab}{\textsc{Barabasi-Albert}\xspace}
\newcommand{\barabss}{\textsc{bar}\xspace}
\newcommand{\emaild}{\textsc{Email EU}\xspace}
\newcommand{\emaildss}{\textsc{ema}\xspace}

\newcommand{\faceb}{\textsc{facebook}\xspace}
\newcommand{\facebss}{\textsc{fac}\xspace}

\newcommand{\youtub}{\textsc{youtube}\xspace}
\newcommand{\youtubss}{\textsc{ytb}\xspace}

\newcommand{\google}{\textsc{Google}\xspace}
\newcommand{\googless}{\textsc{goo}\xspace}
\newcommand{\ita}{\textsc{Italy}\xspace}
\newcommand{\itass}{\textsc{ita}\xspace}
\newcommand{\amazon}{\textsc{Amazon}\xspace}
\newcommand{\amazonss}{\textsc{amz}\xspace}

\newcommand{\arxiv}{\textsc{ArXiv}\xspace}
\newcommand{\arxivss}{\textsc{arx}\xspace}

\newcommand{\bitcoin}{\textsc{bitcoin}\xspace}
\newcommand{\bitcoinss}{\textsc{btc}\xspace}

\newcommand{\cisco}{\textsc{cisco}\xspace}
\newcommand{\ciscoss}{\textsc{cis}\xspace}

\newcommand{\biolog}{\textsc{Bio-Markers}\xspace}
\newcommand{\biologss}{\textsc{bio}\xspace}

\begin{document}

% %
% \newcommand\relatedversion{}
% \renewcommand\relatedversion{\thanks{\mat{da fare}The full version of the paper can be accessed at \protect\url{https://arxiv.org/abs/0000.00000}}} % Replace URL with link to full paper or comment out this line

\title{On Computing Top-$k$ Simple Shortest Paths from a Single Source}
    \author{%Anonymous Author(s)
Mattia D'Emidio\thanks{Department of Information Engineering, Computer Science and Mathematics, and Centre of Excellence Ex-EMERGE (Centre of EXcellence on Connected, Geo-localized and Cyber-secure vehicles), University of L'Aquila (Italy)  --\href{mattia.demidio@univaq.it}{mattia.demidio@univaq.it}, \url{http://www.mattiademidio.com}.}
\and Gabriele Di Stefano\thanks{Department of Information Engineering, Computer Science and Mathematics, and Centre of Excellence Ex-EMERGE (Centre of EXcellence on Connected, Geo-localized and Cyber-secure vehicles), University of L'Aquila (Italy) -- \href{gabriele.distefano@univaq.it}{gabriele.distefano@univaq.it}, \url{http://gs.ing.univaq.it/}.}
}

\date{}

\maketitle

\begin{abstract}
We investigate the problem of computing the \textit{top-$k$ simple shortest paths} in weighted digraphs.
While the single-pair variant -- finding the top-$k$ simple shortest paths between two specified vertices -- has been extensively studied over the past decades, with Yen's algorithm and its heuristic improvements emerging as the most effective solving strategies, relatively little attention has been devoted to the more general single-source version, where the goal is determining top-$k$ simple shortest paths from a source vertex to all other vertices.
Motivated by the numerous practical applications of ranked shortest paths, in this paper we provide new insights and algorithmic contributions to this problem.
In particular, we first present a theoretical characterization of the structural properties of its solutions. Then, we introduce the first polynomial-time algorithm specifically designed to handle it. On the one hand, we prove our new algorithm is on par, in terms of time complexity, with the best (and only) polynomial-time approach known in the literature to solve the problem, that is applying the fastest single-pair algorithm independently to each vertex pair formed by the source and the remaining vertices.
On the other hand, through an extensive experimental evaluation on both real-world and synthetic graphs, we demonstrate that our algorithm consistently and significantly outperforms the latter baseline in terms of running time, achieving speed-ups of up to several orders of magnitude.
These results establish our new algorithm as the solution to be preferred for computing $k$ simple shortest paths from a single source in practical settings.
\end{abstract}

\section{Introduction}
Computing shortest paths is universally considered one of the most fundamental operations to be performed on graph data, as it serves a broad spectrum of applications in computer science and engineering, including communication protocols \cite{cidon1995greedy}, trip planning~\cite{delling2014robust}, vehicle routing~\cite{archetti2013optimal}, network design~\cite{BoklerC20}, timetabling~\cite{Jacob2008} and artificial reasoning~\cite{AnirbanW0K0H22}. 
The \textit{$k$ shortest paths} problem is a natural and long-studied generalization of the problem of determining shortest paths in which the goal is to compute not just a single path but a collection of so-called \textit{top-$k$ shortest paths}, in non-decreasing order of weight~\cite{Yen71,Eppstein98,ZoobiCN23}.
Compared to classical shortest paths, this generalization supports more complex applications, particularly all those that require determining and evaluating, according to some metric, multiple connection options between the vertices of a graph such as, e.g., adaptive motion planning, secure communications in multi-hop networks, fraud detection, natural language processing~\cite{LiXLHZ023,Luo22,AkibaHNIY15,ZoobiCN23,DAscenzoD24,LiuJYZ18a,Eppstein98}.
Moreover, top-$k$ shortest paths are considered a more informative measure of distance between vertices and, by extension, between graphs. Consequently, they are widely used for graph learning tasks~\cite{Eppstein98}, link prediction and social network analytics~\cite{DD23,AkibaHNIY15}.

Computationally speaking, on the one hand the problem of determining $k$ shortest paths presents similarities with the computation of standard shortest paths; in fact, variants of the problem based on the input graph being directed or not (weighted or not, respectively)~\cite{katoh1982efficient,RZ12} and/or on the scope of the computation, namely whether the objective is to compute paths for a pair of vertices (\textit{single-pair} variant)~\cite{Eppstein98,ZoobiCN23,KM16,Gao10}, from a single source to all other vertices (\textit{single-source} variant)~\cite{Eppstein98}, or for all pairs of vertices (\textit{all-pairs} variant)~\cite{AgarwalR16}, have been defined and studied.
On the other hand, finding $k$ shortest paths presents additional algorithmic challenges, primarily due to structural properties of the sought output. In fact, while a shortest path is, by definition, \textit{simple} (i.e. it does not contain \textit{loops}), a collection of $k$ shortest paths may include non-simple paths~\cite{Eppstein98}. Furthermore, $k$ shortest paths lack of optimal sub-structure, and this precludes the possibility of efficiently constructing solutions to problem instances by combining solutions to sub-problems~\cite{AgarwalR16,AkibaHNIY15}.

When paths are not constrained to be simple, the reference approach is Eppstein's algorithm, which computes $k$ shortest paths for a single vertex pair in $\bigO(m + n \log n + k)$ time on any $n$-vertex $m$-edge weighted graph~\cite{Eppstein98}.
This algorithm can be improved to run in $\bigO(n+m+k)$ time when the managed graph is unweighted, and it can be extended to find the $k$ shortest paths from a specified source vertex to all other vertices, or for all vertex pairs,
in $\bigO(m + n \log n + kn)$ and $\bigO(mn + n^2 \log n + kn^2)$ time, respectively~\cite{Eppstein98}.

When paths are required to be simple, the problem becomes more computationally challenging~\cite{AgarwalR16}.
For the single-pair version (commonly referred to as the \textit{\spksisp} problem, or \spksispshort for short), the most efficient known algorithm for general graphs remains the method designed by Yen in 1971~\cite{Yen71}, which computes $k$ simple shortest paths between two vertices in $\bigO(kn(m + n \log n))$ time. 
Despite extensive research over the past decades~\cite{Gao10,KM16,Feng14,ZoobiCN23,PASCOAL_dags,GOTTHILF2009352,FengCLJ23,num_k_paths}, no asymptotic improvement over Yen's result has been achieved for general graphs, while some progress has been made either for special graph classes or from a practical viewpoint.
Concerning special graph classes, for example, Pascoal et al.~\cite{PASCOAL_dags} showed that in directed acyclic graphs (DAGs), $k$ shortest paths can be computed in $\bigO(m+k(n+\log k))$ time using $\bigO(k+m)$ additional space.  If the DAG has also integer edge weights, the time complexity improves to $\bigO(m + kn)$, with the same space requirements. Similarly, Gotthilf at al.~\cite{GOTTHILF2009352} slightly improved Yen's bound for sparse graphs, achieving a complexity of $\bigO(k(mn + n^2 \log \log n))$ while Katoh et al. designed a method that solves \spksispshort in $\bigO(k \cdot c(n, m))$ time in undirected non-negatively-weighted graphs, where $c(n, m) \in \Omega(m)$ is the time complexity of any algorithm for single-source shortest paths~\cite{katoh1982efficient}. 
In addition, Coudert et al.~\cite{COUDERT25} showed that 
$k$ shortest simple distances (paths, respectively) can be computed in $\bigO(k+n)$ time ($\bigO(kn)$ time, respectively) on graphs with treewidth at most $2$ (on bounded treewidth graphs, respectively). 
Finally, from an inherent complexity viewpoint, Gotthilf at al.~\cite{GOTTHILF2009352} showed that \spksispshort is not asymptotically harder than the all-pairs shortest paths (APSP) problem, as it can be solved through $\bigO(k)$ iterations of any algorithm for APSP. Moreover, Vassilevska et al.~\cite{WilliamsW10} proved that any sub-cubic algorithm for \spksispshort would imply a sub-cubic algorithm for APSP.
Concerning empirical progresses, approaches worth to be mentioned are the \emph{node-classification} algorithm, proposed independently in~\cite{Feng14} and~\cite{Gao10}, the \emph{sidetrack-based} method by Kurz and Mutzel~\cite{KM16}, the \textit{postponed node classification} (\pnc) algorithm in~\cite{ZoobiCN23} and the very recent \textit{PeeK} framework~\cite{FengCLJ23}. 
Although all these approaches match Yen's time complexity, they have been shown, through experimentation, to run faster than Yen's for several, practically relevant combinations of graph types and values of $k$ (e.g. road networks for $k$ in the order of hundreds)~\cite{ZoobiCN23,Gao10}, with \pnc emerging as the method that provides the largest speed-ups in the majority of the tested cases, without increasing the space occupancy. Instead, the performance gains of other algorithms come at the price of a high memory overhead (see~\cite{KM16,ZoobiCN23}) or via parallelism~\cite{FengCLJ23}.

In contrast to the non-simple case and to classical shortest paths, both the single-source and the all-pair versions of the problem of computing $k$ \textit{simple} shortest paths have only marginally been investigated~\cite{ShihP12,AgarwalR16}, notwithstanding numerous real-world applications rely on solutions to such problems~\cite{yuan2025service,DBLP:conf/edbt/ChangLQYP15,fi17060232,jariyasunant2011mobile,chen2025lightweight,SHIH2023103760,nitheesh2025disaster,b306,LiuJYZ18a}.
In particular, for the single-source variant (called the \textit{\ssksisp} problem or \ssksispshort for short), to the best of our knowledge, no \textit{dedicated} polynomial-time algorithm is currently available in the literature for general graphs and $k$. The only polynomial-time approach is, straightforwardly, to execute any polynomial-time algorithm for \spksispshort independently for each of the $n-1$ vertex pairs formed by the source and every other vertex of the graph. 
For the all-pairs variant, instead, a dedicated solving approach is that of Argawal et al.~\cite{AgarwalR16}, which is asymptotically better than computing $k$ shortest paths for all vertex pairs by $\Theta(n^2)$ executions of Yen's algorithm, but works only under the very restrictive hypothesis of $k\in\{2,3\}$. 
Similarly to \ssksispshort, the only polynomial-time solving strategy to handle the all-pairs variant for general $k$ is to repeat any polynomial-time algorithm for \spksispshort for all $\binom{n}{2}$ vertex pairs.
However, empirical evidence~\cite{ZoobiCN23,Gao10,FengCLJ23} suggests that adopting the above-mentioned approaches induces prohibitively long running times even for small graphs and values of $k$, rendering it impractical for the application domains in which $k$ simple shortest paths for many pairs must be computed routinely.
% %
\blocco{Our Contribution}
In this paper, we take a step forward in addressing these practical limitations by introducing a novel algorithm to solve \ssksispshort on general digraphs. Our approach is based on a theoretical characterization of the structural properties of the solutions to the problem. We prove the correctness of the algorithm and analyze its worst-case time complexity, showing that it runs in polynomial time with respect to both the graph size and $k$.
On the one hand, this is on par, in terms of asymptotical time complexity, with the state-of-the-art option to attack the problem, that is running the most efficient, polynomial-time algorithm for \spksispshort $n-1$ times from the source to every other vertex.
On the other hand, we provide the results of an extensive experimental evaluation, on both real-world and synthetic datasets, that demonstrates that our new algorithm consistently outperforms this state-of-the-art method, since, in all cases, it computes solutions to \ssksispshort significantly faster, often by orders of magnitude.

\section{Preliminaries}
\label{sec:prelim}
In this section, we give the notation and the definitions used throughout the paper.
% %
We are given a directed non-negatively weighted graph $G=(V,E,\omega)$, having $n=|V|$ vertices and $m=|E|$ edges, with $E \subseteq V \times V$ and $\omega:E\rightarrow \mathbb{R}^{+}$.
In the remainder of the paper, we use terms graph and digraph interchangeably. 
We use $\nout(v)=\{u \in V | (v,u) \in E\}$ ($\nin(v)=\{u \in V | (u,v) \in E\}$, respectively) to denote the set of \textit{outgoing} (\textit{incoming}, respectively) neighbors of a vertex $v\in V$. A \textit{path} $P=(s\equiv v_1,v_2,\dots,t\equiv v_\eta)$ in $G$, connecting a pair of vertices $s,t \in V$ (called its \textit{endpoints}), is a sequence of $\eta$ vertices such that $(v_i,v_{i+1})\in E$ for all $i\in[\eta-1]$.
Given an integer $k>0$, we let $[k]:=\{1,2,\ldots,k\}$.
%Moreover, 
Given a path $P=(v_1,v_2,\dots,v_\eta)$ in $G$, we denote by $V(P)$ ($E(P)$, respectively) the vertices (edges, respectively) of $P$. Moreover, we say a path $P$ is \emph{simple} (or \emph{loopless}) when there are no vertex repetitions in $V(P)$. Finally, we call $\wg{P}=\sum_{i=1}^{\eta-1} \omega(v_i,v_{i+1})$ the \textit{weight} of a path $P$.
%the sum of the weights of the edges induced by the vertices in $P$, i.e. 
%
In an unweighted graph, the weight is equal to the \textit{length} of the path, that is, to the number of its edges. 
%(unweighted graphs can be treated as weighted ones with unitary weights).
% Note that, for non-simple paths, the weight of a path includes multiple occurrences of a same edge, if any.
Given a graph $G=(V,E,\omega)$ and a subset $V'\subseteq V$ of its vertices, we denote by $G[V']$ the sub-graph of $G$ induced by vertices in $V'$, i.e. graph $G'=(V',E',\omega)$ where $E'=\left\{(u,v) \in E\,|\,u,v \in V'\right\}$. 

A \textit{shortest path}, for two vertices $s,t \in V$, is a path having minimum weight among all paths connecting $s$ to $t$.
% The \textit{distance} $d(s,t)$ between $s$ and $t$ is the weight \wg{P} of a shortest path $P$ from $s$ to $t$. 
%
Given two paths, say $P'=(v_1,v_2,\dots,v_\eta)$ and $P''=(u_1,u_2,\dots,u_\zeta)$ such that $v_{\eta}\equiv u_1$, we denote by $P'\oplus P''$ the \textit{concatenation} of the two paths, that is, path $(v_1,v_2,\dots,v_\eta\equiv u_1,u_2,\dots,u_\zeta)$ obtained by appending all vertices of $P''$ to $P'$. If $P=P'\oplus Q$ (with $Q$ possibly being the empty path), we call $P'$ a \emph{prefix} of $P$.

Given a graph $G=(V,E,\omega)$ with $\omega:E\rightarrow \mathbb{R}^{+}$, two vertices $s,t \in V$, and an integer $k > 0$, the \textit{\spksisp} problem (\spksispshort, for short)  asks to compute a collection $\soluzionepair{s}{t}=\{P_1,P_2,\dots, P_{k'}\}$ of $k'\leq k$ paths from $s$ to $t$ such that:
(i) $P_i$ is simple for any $i\in[k']$; (ii) $P_i\neq P_j$ for any $i,j\in [k'], i \neq j$; (iii) $\wg{P'}\geq \wg{P_i}$ for any path $P'$ from $s$ to $t$ that is not in $\soluzionepair{s}{t}$ and for any $P_i\in \soluzionepair{s}{t}$; (iv) $\soluzionepair{s}{t}$ is maximal (there does not exist a collection of size $k''>k'$). 
By the above, it follows that: if there exist $k'< k$ simple paths from $s$ to $t$ in $G$, then the problem requires the computation of a collection $\soluzionepair{s}{t}$  that contains all and only such $k'<k$ paths; otherwise, the problem asks to compute a collection $\soluzionepair{s}{t}$ of $k$ pair-wisely distinct simple paths from $s$ to $t$ such that the weight \wg{P'} of any path $P'$ that is in $G$ but not in $\soluzionepair{s}{t}$  is larger than or equal to the weight of any path in $\soluzionepair{s}{t}$.
In either case, a collection of paths from $s$ to $t$, satisfying the constraints imposed by \spksispshort, is called a collection of top-$k$ simple shortest paths for pair $(s,t)$. Observe that there can be several feasible collections (e.g., there exist $k'' > k$ paths of same weight for a pair).

A generalization of \spksispshort is the \textit{\ssksisp} problem (\ssksispshort)  which, given a \textit{source} (or \textit{root}) vertex $r \in V$, asks to compute, for each $v \in V \setminus\{r\}$, a collection $\solssvertex{v}=\{P_1,P_2,\dots, P_{k'}\}$ of $k'\leq k$ paths from $r$ to $v$ which is a solution to \spksispshort for pair $(r,v)$. 
Each $\solssvertex{v}$ is called a collection of top-$k$ simple shortest paths from $r$ to $v$ (or for $v$, when the root is clear from the context).
% \par 
As in \spksispshort, for each vertex $v$, there can be several feasible collections (e.g., if there exist $k'' > k$ paths of same weight from the root $r$ to a vertex $v$). 
We call $\collezionesol{v}$ the family of all such collections for a vertex $v$ and use $\solss_r=\{\solssvertex{v_1},\solssvertex{v_2},\ldots,\solssvertex{v_n}\}$ to identify any solution to \ssksispshort, where $\solssvertex{v_i} \in \collezionesol{v_i}\,\ ~\forall~ v_i\in V$. 
We write $\solss$ instead of $\solss_r$  when $r$ is clear from the context. 

\section{Characterizing Solutions to \ssksispshort}
\label{sec:char}
In this section, we first formalize a brute-force approach, named Algorithm~\exhaustive, for solving \ssksispshort and show that its time complexity is exponential with respect to the input size in the worst case. 
Then, we present a characterization of some relevant properties of solutions to \ssksispshort which we leverage to improve Algorithm~\exhaustive and obtain an algorithm that solves the problem in polynomial time.
\subsection{Solving \ssksispshort Exhaustively} 
In what follows we introduce Algorithm~\exhaustive, an algorithm for \ssksispshort obtained by modifying the brute-force method to find all simple paths in a graph for a pair of vertices~\cite{graph_algorithms,AkibaHNIY15}. 
The algorithm (whose pseudo-code is given in Algorithm~\ref{algo:exhaustive}) solves \ssksispshort by exhaustively discovering, through a modified Dijkstra's-like visit, all simple paths from a given root vertex to any other vertex $v\in V$ of the graph and by storing, in a data structure named $\bestk{v}$, only a subset of such paths that form the sought collection of top-$k$ simple shortest paths for $v$. 
To this end, the algorithm: (i) finds, enqueues and dequeues simple paths from the root in a minimum priority queue $PQ$, where the priority is the weight of discovered paths; (ii) terminates when $PQ$ is empty. 
\begin{algorithm2e}[t]
\small
\SetAlgoLined
\KwIn{Graph $G=(V,E,\omega)$, root $r \in V$, $k\in \mathbb{N}$.}
\KwOut{Top-$k$ simple shortest paths $\bestk{v}$ from $r$ to each $v \in V \setminus\{r\}$.}
$PQ\gets \emptyset$\tcp*{Empty priority queue} 
\ForEach(\tcp*[f]{Empty lists}){$v\in V$}{$\bestk{v}  \gets [~]$\;}
$PQ.enqueue((r),\wg{(r)}=0)$\;
\While(\label{line:while}){$PQ\neq \emptyset$}{
 $\Pi=(r,\dots,v) \gets PQ.dequeueMin()$  \label{line:startloop}\;
\ForEach(\tcp*[f]{Simple paths}){$u \in \nout(v) \setminus V(\Pi)$}{
$PQ.enqueue(\Pi \oplus (u),\wg{\Pi}+\omega(v,u))$\label{line:enqueue}\;}
\If{$|\bestk{v}|<k$}{Add $\Pi$ to $\bestk{v}$\label{line:add}\;}
\label{line:endloop}
}
\caption{Algorithm \exhaustive.}
\label{algo:exhaustive}
\end{algorithm2e}
In the description of the procedures, given a collection of paths $\bestk{v}$, we use $V(\bestk{v})$ to denote the set of vertices that belong to paths in $\bestk{v}$, i.e. $V(\bestk{v})=\cup_{P \in \bestk{v}} V(P)$.
%similarly to the notation used for solutions to \ssksispshort. 
Moreover, in the remainder of the paper, we assume that routine $PQ.enqueue(p,\delta)$ adds to $PQ$ an element $p$ with priority $\delta$, while routine $PQ.dequeueMin()$ returns the element of $PQ$ having minimum priority and removes it from $PQ$.

Now, we first introduce and prove two lemmas that are necessary to show the correctness of algorithm~\exhaustive.
\begin{lemma}\label{lemma:exhaustivesearch}
Algorithm~\exhaustive computes all simple paths from the root vertex to any other vertex.
\end{lemma}
\begin{proof}
By contradiction, assume that there exists a simple path from the root $r$ to a vertex that is not found by Algorithm \exhaustive, which means that it is never dequeued from $PQ$ in line~\ref{line:dequeue}. Among all such simple paths, let $\pi=(r,\dots,v,u)$ be the shortest. Since every path enqueued is eventually extracted from $PQ$, the prefix $\pi'=(r,\dots,v)$ must have been both enqueued and dequeued; otherwise, $\pi$ would not be the shortest path that is not found by the algorithm.
It follows that, when $\pi'$ is dequeued, $\pi$ is enqueued in line~\ref{line:enqueue}. Moreover, $\pi$ is simple since $u \notin V(\pi')$. This contradicts our hypothesis, as it implies that $\pi$ is enqueued and eventually dequeued from $PQ$.
\end{proof}

\begin{lemma}\label{lemma:nonincreasing}
Algorithm \exhaustive dequeues simple paths from the root vertex to any other vertex in non-decreasing order of weight.
\end{lemma}
\begin{proof}
Assume by contradiction that there exist two paths, say $\pi$ and $\pi'=(r,\dots,v)$, such that $\wg{\pi}>\wg{\pi'}$ and $\pi$ is dequeued from $PQ$ before $\pi'$.
Call $F$ the set of paths that are dequeued by the algorithm before $\pi$ is dequeued. Observe that $F$ contains at least path $(r)$ which is enqueued and dequeued in any case by the algorithm in its first steps. Now, focus on when path $\pi$ is dequeued.
Since $\wg{\pi}>\wg{\pi'}$ we have $\pi'\not\in PQ$, as otherwise we would have dequeued $\pi'$. since the algorithm extracts the minimum. 
Moreover, $\pi'\notin F$ as well by hypothesis.
Hence, the prefix of $\pi'$, say $\pi''$, such that $\pi'=\pi''\oplus (v)$, cannot be in $PQ$ and neither in $F$. In fact, $\pi''$ cannot be in $PQ$, as otherwise $\pi''$ would have been dequeued in place of $\pi$ (note that $\pi''$ is shorter than $\pi$ since $\wg{\pi'}<\wg{\pi}$) and therefore $\pi'$ would have been enqueued in line~\ref{line:enqueue} and therefore extracted before $\pi$, which is a contradiction. Moreover, $\pi''$ cannot be in $F$, as otherwise $\pi'$ would have been added to $PQ$ when $\pi''$ is dequeued.
Recursively repeating this argument for any prefix of $\pi'$, we reach the contradiction of the path $(r)$ not being in both $PQ$ and $F$.
\end{proof}
The correctness of Algorithm~\ref{algo:exhaustive} is shown by the following results.

\begin{theorem}
Algorithm~\ref{algo:exhaustive} solves \ssksispshort.
\label{th:exhaustive:correct}
\end{theorem}
\begin{proof}
The proof follows by Lemmas~\ref{lemma:exhaustivesearch} and~\ref{lemma:nonincreasing}.
In particular, by Lemma~\ref{lemma:nonincreasing} we know that Algorithm \exhaustive discovers all simple paths in order of weight. Hence, in line~\ref{line:add}, data structure \bestk{v} is filled with paths that follow a non-decreasing order of weight, for each $v\in V$. Moreover, by Lemma~\ref{lemma:exhaustivesearch} we know that Algorithm \exhaustive determines all simple paths from the root, hence at termination $\bestk{v}=\solssvertex{v}$ for each $v\in V$ and for some $\solssvertex{v}\in \collezionesol{v}$.
\end{proof}
We emphasize that all methods introduced in this paper are designed for directed weighted graphs but can be applied to both undirected or unweighted graphs by considering bidirectional edges (i.e. $\nout(v)=\nin(v)$ for each $v \in V$) or unitary weights (i.e. $\wg{u,v}=1$ for each $(u,v)\in E$).
Moreover, note that Algorithm~\ref{algo:exhaustive} can be modified to terminate earlier by adding a conditional statement inside the while loop of line~\ref{line:while}. Specifically, if we define a vertex $v$ to be \textit{saturated} when $|\bestk{v}|=k$, the loop can be stopped when all vertices are saturated.
It is easy to see that this modification does not affect the algorithm's correctness (since simple paths are found in non-decreasing order of weight) nor its time complexity (in the worst case, there may be a vertex $w$ for which fewer than $k$ simple paths from $r$ to $w$ exist in the graph).
In the remainder of this paper, with a little abuse of notation, we refer to this variant as Algorithm~\exhaustive as well.
Note also that, while Algorithm~\ref{algo:exhaustive} is simple and elegant, its time complexity is exponential in the input size, as shown below:
\begin{theorem}
There exists an $n$-vertex graph that forces algorithm \exhaustive to run for $\Omega(2^n)$ time to achieve a solution to \ssksispshort.
\label{th:exhaustive:runningtime}
\end{theorem}
\begin{proof}
% In the worst case, the algorithm adds all simple paths from the root to any vertex $v\in V$ into $PQ$. 
% We construct an input instance where the number of paths added to $PQ$, before extracting a shortest path from the root to a vertex, grows exponentially w.r.t. $n$, regardless of $k$. 
Consider the undirected, unweighted graph in Fig.~\ref{fig:exhaustive:exponential}: call $d$ the number of vertices of type $c_i$. Then, the number of paths that are enqueued in $PQ$, before extracting a shortest path from the root $r$ to $v$, is at least the number of simple paths in the graph from $r$ to $c_d$, that is, $2^{d}$. Since $d=\Theta(n)$ the claim follows.

\end{proof}
\begin{figure}[t]
\centering
\subfloat[\centering]{
\resizebox{.49\textwidth}{!}{ 
        \begin{tikzpicture}[node/.style={circle, draw=black, very thick, minimum size=7mm, scale=.75}]
            %% Vertices
            \node[node] (r) {$r$};
            % \node[node]  (x1)     [right=.5cm of r] {};
            \node[node]  (x1)     [above right=.3cm of r] {};
            \node[node]  (x2)     [below right=.3cm of r] {};
            \node[node]  (y1)     [below right=.3cm of x1] {$c_1$};
            \node[node]  (z1)     [above right=.3cm of y1] {};
            \node[node]  (z2)     [below right=.3cm of y1] {};
            \node[node]  (w1)     [below right=.3cm of z1] {$c_2$};
            \node[node]  (e1)     [right=1.5cm of w1] {$c_d$};
            \node[node]  (s1)     [above right=.3cm of e1] {};
            \node[node]  (s2)     [below right=.3cm of e1] {};
            \node[node]  (v)     [below right=.3cm of s1] {$v$};
            %% Edges
            \draw[-, line width=.2mm] (r.north east) -- (x1);
            \draw[-, line width=.2mm] (r.south east) -- (x2);
            \draw[-, line width=.2mm] (x1) -- (y1);
            \draw[-, line width=.2mm] (x2) -- (y1);
            \draw[-, line width=.2mm] (y1.north east) -- (z1);
            \draw[-, line width=.2mm] (y1.south east) -- (z2);
            \draw[-, line width=.2mm] (z1) -- (w1);
            \draw[-, line width=.2mm] (z2) -- (w1);
            \draw[] ($(w1)!0.35!(e1)$) edge [dashed, very thick] ($(w1)!0.61!(e1)$);
            \draw[-, line width=.2mm] (e1.north east) -- (s1);
            \draw[-, line width=.2mm] (e1.south east) -- (s2);
            \draw[-, line width=.2mm] (s1) -- (v);
            \draw[-, line width=.2mm] (s2) -- (v);
        \end{tikzpicture}
        }
        \label{fig:exhaustive:exponential}
}
\subfloat[\centering]{
\resizebox{.49\textwidth}{!}{ 
\begin{tikzpicture}[node/.style={circle, draw=black, very thick, minimum size=7mm, scale=.75}]
            %% Vertices
            \node[node] (r) {$r$};
            \node[node]  (x1)     [above right=.3cm of r] {$x_1$};
            \node[node]  (x2)     [below right=.3cm of r] {$x_2$};
            \node[node]  (y1)     [below right=.3cm of x1] {$c_1$};
            \node[node]  (z1)     [above right=.3cm of y1] {$x_3$};
            \node[node]  (z2)     [below right=.3cm of y1] {$x_4$};
            \node[node]  (w1)     [below right=.3cm of z1] {$c_2$};
            \node[node]  (e1)     [right=1.2cm of w1] {$c_d$};
            \node[node]  (s1)     [above right=.3cm of e1] {};
            \node[node]  (s2)     [below right=.3cm of e1] {};
            \node[node]  (v)     [below right=.3cm of s1] {$v$};
            \node[node]  (o1)     [above right=.3cm of x1] {};
            \node[node]  (o2)     [above left=.3cm of s1] {};
            %% Edges
            \draw[-, thick] (r.north east) -- (x1);
            \draw[-, thick] (r.south east) -- (x2);
            \draw[-, thick] (x1) -- (y1);
            \draw[-, thick] (x2) -- (y1);
            \draw[-, thick] (y1.north east) -- (z1);
            \draw[-, thick] (y1.south east) -- (z2);
            \draw[-, thick] (z1) -- (w1);
            \draw[-, thick] (z2) -- (w1);
            \draw[] ($(w1)!0.35!(e1)$) edge [dashed, very thick] ($(w1)!0.65!(e1)$);
            \draw[-, thick] (e1.north east) -- (s1);
            \draw[-, thick] (e1.south east) -- (s2);
            \draw[-, thick] (s1) -- (v);
            \draw[-, thick] (s2) -- (v);
            \draw[red, thick] (x1) -- (o1);
            \draw[dashed, red, thick] (v.north) to [out=90,in=30] (o2.east); %(v.north) -- (o2.east);
            \draw[dashed,red, thick]   (o1) edge (o2);
        \end{tikzpicture}
        \label{fig:basic:exponential}
        }
        }
\caption{\protect\subref{fig:exhaustive:exponential} and \protect\subref{fig:basic:exponential}: input instances considered in the proofs of Theorems~\ref{th:exhaustive:runningtime} and~\ref{th:basic:runningtime}, respectively.}
\end{figure}
We now provide a characterization of some important properties of solutions to \ssksispshort to exploit to design efficient algorithms to solve the problem.
In particular, we are interested at identifying conditions that can be tested to stop early the search of algorithm \exhaustive, to reduce its time complexity.
To this aim, we first provide the following definition.
\begin{definition}[\sc Predecessor]
Let $\solss=\{\solssvertex{v_1},\solssvertex{v_2},\ldots,\solssvertex{v_n}\}$ be a solution to \ssksispshort  for a given graph $G=(V,E,\omega)$, root vertex $r\in V$, and $k>0$.
%x->u
%y->v
Consider the relation $R_\solss\subseteq V\times V$ such that $u R_\solss v$ if and only if vertex $u \in V\setminus\{v\}$ belongs to a path of $\solssvertex{v}$ for $v\in V$, i.e. $u\in V(P)\setminus\{v\}$ for some $P\in \solssvertex{v}$.
Then, we say vertex $u$ is a \textit{predecessor} of $v$ in $S$.
\label{def:predecessor}
\end{definition}
We denote by $V(\solssvertex{v})=\{u~|~u R_\solss v\}$ a set of \textit{predecessors} of a vertex $v$.
The next definition introduces the notion of \textit{general predecessor} of a vertex in a solution $S$ to \ssksispshort. Informally, $v$ is a general predecessor of itself and, if a vertex $u$ is a general predecessor of $v$, then the predecessors of $u$ are also general predecessors of $v$. Formally:
\begin{definition}[\sc General Predecessor]
\label{def:general_predecessor}
Let $\solss$ be a solution to \ssksispshort for a given graph $G=(V,E,\omega)$, root vertex $r\in V$, and $k>0$.
Let $T_\solss$ be the transitive closure
%\footnote{We refer the reader to \cite{Grami20221} for a definition of transitive closure of a relation.} 
of relation $R_\solss$ of Def.~\ref{def:predecessor}. 
A vertex $u \in V$ is a \textit{general predecessor} of a vertex $v \in V$  if and only if $u T_\solss v$ or $u=v$. 
\end{definition}
We denote by $A_{\solss,v}=\{u~|~u T_\solss v\}\cup\{v\}$ a set of \textit{general predecessors} of a vertex $v \in V$.
The following lemma shows that sets of general predecessors form a nested set collection. 
\begin{lemma}
\label{lem:recursive}
Let \solss be a solution to \ssksispshort for a given graph $G=(V,E,\omega)$, root vertex $r\in V$, and $k>0$.
Then, for any $v,x\in V$ such that $x \in A_{\solss,v}$, we have $A_{\solss,x} \subseteq A_{\solss,v}$.
\end{lemma}
\begin{proof}
By Definition~\ref{def:general_predecessor}, any vertex $y \in A_{\solss,x}$ is a general predecessor of $x$. Hence $y$ is such that $y T_\solss x$. Similarly, since $x \in A_{\solss,v}$, we have $x T_\solss v$. By transitivity, it follows that $y T_\solss v$ and hence that $y \in A_{\solss,v}$.
\end{proof}
The following two lemmata establish important structural properties that hold for certain paths belonging to solutions to \ssksispshort.
In particular, Lemma~\ref{lem:basic} shows that if there exists a path $P_{rv}$ that \textbf{is not in any }solution, for some vertex $v\in V$, and such that $P_{rv}$ is a prefix of a path $P_{rw}$ that is in \textbf{all} solutions for some other vertex $w\in V$, then a vertex of $P_{rw}$ must be a predecessor of $v$.
Figure~\ref{fig:lemma:basic} provides a visual representation of the claim of Lemma~\ref{lem:basic}.
\begin{lemma}[\sc Basic Lemma]
\label{lem:basic}
Let $\solss=\{\solssvertex{v_1},\solssvertex{v_2},\ldots,\solssvertex{v_n}\}$ be a solution to \ssksispshort for a given graph $G=(V,E,\omega)$, root vertex $r\in V$, and $k>0$.
Let $P_{rw}=P_{rv}\oplus P_{vw}$ any path from the root $r$ to some vertex $w \in V$ such that $P_{rw} \in \solssvertex{w}$, $\forall\, \solssvertex{w}\in \collezionesol{w}$ and $P_{rv}\not \in \solssvertex{v}$.
Then, there exists a vertex $t\in V(\solssvertex{v})\cap V(P_{vw})$.
\end{lemma}
\begin{proof}
Notice that, by the definition of $V(\solssvertex{v})$, we have $t\not =v$ while $t$ might coincide with $w$ or not.
By contradiction, assume that $ V(\solssvertex{v})\cap V(P_{vw})$ is empty. Consider the set of simple paths $\Pi_{rw}=\{Q\oplus P_{vw}~|~Q\in \solssvertex{v}\}$. Since the number of paths in $\solssvertex{v}$ is $k$ -- otherwise $P_{rv}$ should belong to $\solssvertex{v}$ -- also the size of $\Pi_{rw}$ is $k$. Furthermore, the weight of each path in $\Pi_{rw}$ is less than or equal to $\wg{P_{rw}}$. Hence, there exists a solution $\solss'$ to \ssksispshort such that $P_{rw}\not \in \solssvertex{w}'$, a contradiction.
\end{proof}
A stronger characterization of the structure of paths that belong to solutions to \ssksispshort is provided through the following result.
The next lemma asserts that if a path $P_{rw}=P_{rv}\oplus P_{vw}$ is necessarily in a solution for a vertex $w$, but the prefix $P_{rv}$ is not necessary to complete a solution for $v$, then there must be a vertex $y$ in the general predecessors of $v$ that lies on $P_{vw}$ and such that the prefix $P_{ry}$ of $P_{rw}$ is among the paths in a solution for $y$.
Figure~\ref{fig:lemma:path} provides a visual representation of the statement of Lemma~\ref{lem:path}.
\begin{figure}[t]
\centering
\subfloat[\centering]{
  \scalebox{.75}{
  	\begin{tikzpicture}[font=\large,semithick,transform shape,minimum size=6mm,inner sep=0pt]
       \node[draw,circle,fill=white] (root) at (2,9) {$r$};
        \begin{scope}
         \clip[draw] (root) -- (0,6) -- (2,2) -- (4,6) -- (root);
      \draw[rotate=135,step=0.3cm,blue,thin] (-10,-10) grid (10,10);  
        \end{scope}
        \node[draw,circle,fill=white] (v) at (2,2) {$v$};
        \node[draw,rectangle,fill=white] (M1) at (2,6) {$V(\solssvertex{v})$};
        \node[draw,circle,fill=white] (w) at (5,7) {$w$};
        \node[circle,fill=none] (l1) at (0,6) {};
        \node[circle,fill=none] (l2) at (1.4,3.5) {};

        \node[circle,fill=none] (y) at (4,6) {};
        \node[draw,circle,fill=white] (z) at (3.3,4.5) {$t$};
        \draw[dashed,very thick,->]   (y.center) edge[out=45,in=-135] (w);
        \draw[dashed,very thick,->]   (v) edge[out=0,in=0] (z);
        \draw[dashed,very thick,->]   (z) edge[out=180,in=180] (y.center);
        \draw[very thick]   (root) edge [out=180,in=180] (l1.center);
        \draw[very thick]   (l1.center) edge[out=0,in=0] (l2.center);

        \draw[very thick,->]   (l2.center) edge [out=180,in=180] (v);
		\end{tikzpicture}
  }
  \label{fig:lemma:basic}
  }
  \subfloat[\centering]{
  \scalebox{.75}{
    	\begin{tikzpicture}[font=\large,semithick,transform shape,minimum size=6mm,inner sep=0pt]
       \node[draw,circle,fill=white] (root) at (2,9) {$r$};
        \begin{scope}
         \clip[draw] (root) -- (1.5,6) -- (3.25,4.5) -- (4,6) -- (root);
        \draw[rotate=135,step=0.3cm,blue,very thick] (-10,-10) grid (10,10);  
        \end{scope}
         \begin{scope}
         \clip[draw] (root) -- (0,6) -- (2,2) -- (4,6) -- (root);
         % \draw[red];
        \draw[rotate=135,step=0.3cm,red,dashed,thin] (-10,-10) grid (10,10);  
        \end{scope} 
        \node[draw,circle,fill=white] (v) at (2,2) {$v$};
        \node[draw,rectangle,fill=white] (M1) at (1.5,6) {$A_{\solss,v}$};
        \node[draw,rectangle,fill=white] (M2) at (2.8,6.5) {$V(\solssvertex{t})$};

        \node[draw,circle,fill=white] (w) at (5,7) {$w$};
        \node[circle,fill=none] (l1) at (0,6) {};
        \node[circle,fill=none] (l2) at (0.9,4.5) {};

        \node[draw,circle,fill=white] (y) at (4,6) {$y$};
        \node[draw,circle,fill=white] (t) at (3.3,4.5) {$t$};
        \draw[red,very thick,->]   (y) edge[out=45,in=-135] (w);
        \draw[dashed,very thick,->]   (v) edge[out=0,in=0] (t);
        \draw[dashed,very thick,->]   (t) edge[out=-135,in=-165] (y);
        \draw[very thick]   (root) edge [out=180,in=180] (l1.center);
        \draw[very thick]   (l1.center) edge[out=0,in=0] (l2.center);
        \draw[very thick,->]   (l2.center) edge [out=180,in=180] (v);
		\end{tikzpicture}     
}
\label{fig:lemma:path}
}
\caption{\protect\subref{fig:lemma:basic}: Structure of solution \solss of Lemma~\ref{lem:basic}: the solid path is $P_{rv}\not\in \solssvertex{v}$ while the dashed path is $P_{vw}$; vertex $t$ must be a predecessor of $v$ in \solss; the blue grid highlights set $V(\solssvertex{v})$. \protect\subref{fig:lemma:path}: Structure of solution $\solss$ of Lemma~\ref{lem:path}: the black solid path is $P_{rv}$ while the black dashed one is $P_{vy}$; the red path is a simple path $P_{yw}$ in $G[V\setminus A_{\solss,v}\cup\{y\}]$; vertices $t,y$ are in $V(P_{vw})\cap A_{\solss,v}$; the blue grid identifies the vertices of $V(\solssvertex{t})$ while the red dashed grid indicates set $A_{\solss,v}$, which includes $V(\solssvertex{t})$.
}
\end{figure}
\begin{lemma}[\sc Path Lemma]
\label{lem:path}
Let $\solss=\{\solssvertex{v_1},\solssvertex{v_2},\ldots,\solssvertex{v_n}\}$ be a solution to \ssksispshort for a given graph $G=(V,E,\omega)$, root vertex $r\in V$, and $k>0$.
Let $P_{rw}=P_{rv}\oplus P_{vw}$ be a path from $r$ to a vertex $w \in V$ such that $P_{rw}\in \solssvertex{w}, \forall\, \solssvertex{w}\in \collezionesol{w}$, and $P_{rv}\not \in \solssvertex{v}$. 
Then, there exist two vertices $t,y$ in $V(P_{vw})\cap A_{\solss,v}$ such that: i) $y \in V(\solssvertex{t})$; ii) $P_{rt}=P_{rv}\oplus P_{vt} \not \in \solssvertex{t}$;  and  iii) $P_{ry}=P_{rv}\oplus P_{vy}  \in \solssvertex{y}$, where $P_{vt}$ and  $P_{vy}$ are both prefixes of $P_{vw}$.
\end{lemma}
\begin{proof}
By Lemma~\ref{lem:basic}, there is a vertex $t_0$ in $P_{vw}$ that belongs to $ V(\solssvertex{v})$.
If $P_{rt_0}=P_{rv}\oplus P_{vt_0} \in \solssvertex{t_0}$ then we are done, since $v$ and $t_0$ play the roles of $t$ and $y$, respectively. On the contrary, we can consider, recursively, vertex $t_0$ and path $P_{rt_0}\not \in \solssvertex{t_0}$ and apply again Lemma~\ref{lem:basic}. Hence, there exists a vertex $t_1$ in $P_{vw}$ belonging to $V(\solssvertex{t_0})$.
Now, if $P_{rt_1}=P_{rt_0}\oplus P_{t_0t_1} \in \solssvertex{t_1}$, again we are done as $t_0$ and $t_1$ play the roles of $t$ and $y$, respectively. Vice versa, we can repeat the reasoning above and find either two internal vertices $t_i$ and $t_{i+1}$ of $P_{rw}$ that play the roles of $t$ and $y$, or an internal vertex $t_i$ of $P_{rw}$ and $w$ itself playing the roles of $t$ and $y$, respectively, since $P_{rw}\in \solssvertex{w}, \forall\, \solssvertex{w}\in \collezionesol{w}$ by hypothesis.
\end{proof}
%%%%%
Finally, we derive another relation between the paths in any solution to \ssksispshort and the general predecessors of the vertices on such paths. When searching a graph for a solution to \ssksispshort, this relation is helpful to safely discard certain paths, as they certainly are not part of any solution, based on conditions holding on their prefixes.
\begin{lemma}[\sc Pruning Lemma]
\label{lem:pruning}
Let $\solss=\{\solssvertex{v_1},\solssvertex{v_2},\ldots,\solssvertex{v_n}\}$ be a solution to \ssksispshort for a given  graph $G=(V,E,\omega)$, root vertex $r\in V$, and $k>0$.
Let $P_{rv}$ any path from $r$ to a vertex $v \in V$. If (i) $|\solssvertex{u}|=k$ for all vertices $u\in A_{\solss,v}\setminus\{r\}$ and (ii) $P_{rv}$ is not a prefix of any path in $\bigcup_{u\in A_{\solss,v}}\solssvertex{u}$, then there exists a solution $\solss'=\{\solssvertex{v_1}',\solssvertex{v_2}',\ldots,\solssvertex{v_n}'\}$ such that $P_{rv}$ is not a prefix of any path in $\bigcup_{u\in V}\solssvertex{u}'$.
\end{lemma}
\begin{proof}
We construct a solution $\solss'=\{\solssvertex{v_1}',\solssvertex{v_2}',\ldots,\solssvertex{v_n}'\}$ by modifying $\solss$ as follows.
By hypothesis, $P_{rv}$ is not a prefix of any path in $\bigcup_{u\in A_{\solss,v}}\solssvertex{u}$ hence we assign $\solssvertex{u}'= \solssvertex{u}$ for all vertices $u\in A_{\solss,v}$.
Now, consider any vertex $w \in V\setminus A_{\solss,v}$ and let $P_{rw}=P_{rv}\oplus P_{vw}\in \solssvertex{w}$ be a path having $P_{rv}$ as prefix and such that $P_{rv}$ satisfies properties (i) and (ii). %
We now show that $P_{rw}$ can be replaced with another path that does not have $P_{rv}$ as prefix.
Let $y$ be the vertex that lies on $P_{vw}$, belongs to set $A_{\solss,v}$ and is closest to $w$, that is a vertex such that $P_{vw}=P_{vy}\oplus P_{yw}$ where $P_{yw}$ is a path from $y$ to $w$ in $G[V\setminus A_{\solss,v}\cup\{y\}]$. Note that $y$ can be $v$ itself.
Let $P^{max}_{ry}$ be the heaviest path in $\solssvertex{y}$. Since $P_{rv}$ is not a prefix of any path in $\bigcup_{u\in A_{\solss,v}}\solssvertex{u}$, it follows that the weight of path $P_{ry}=P_{rv}\oplus P_{vy}$ must be at least that of $P^{max}_{ry}$.
Hence, we have:
\begin{align*}
\wg{P_{rw}}&=\wg{P_{rv} \oplus P_{vw}}=\wg{P_{rv}\oplus P_{vy}\oplus P_{yw}}\\&=\wg{P_{rv}\oplus P_{vy}} + \wg{P_{yw}} \\&\geq \wg{P_{ry}^{max}}+\wg{P_{yw}} \geq \wg{P_{ry}^{max}\oplus P_{yw}}.
\end{align*}
Since $y \in A_{\solss,v}$, we have $|\solssvertex{y}|=k$ by hypothesis $(i)$. Moreover, by Lemma~\ref{lem:recursive}, $A_{\solss,y}\subseteq A_{\solss,v}$. Then, there must exist, in $G$, $k$ simple paths in the form $P_{ry}\oplus P_{yw}$, one for each path $P_{ry} \in \solssvertex{y}$, that all have weight at most  $\wg{P_{rw}}$.
Thus, we can construct set $\solssvertex{w}'$ by replacing $P_{rw}\in \solssvertex{w}$ with a path $P_{ry}\oplus P_{yw}$ for some $P_{ry} \in \solssvertex{y}$.
\end{proof}

\section{New Algorithms for \protect\ssksispshort}
In this section, we introduce new algorithms to solve \ssksispshort. 
\begin{algorithm2e}[t]
\small
\SetAlgoLined
\KwIn{Graph $G=(V,E,\omega)$, root $r \in V$, $k\in \mathbb{N}$.}
\KwOut{Top-$k$ simple shortest paths $\bestk{v}$ from $r$ to each $v \in V \setminus\{r\}$.}
$PQ\gets \emptyset$\tcp*{Empty priority queue} 
\ForEach(\tcp*[f]{Empty lists}){$v\in V$}{$\bestk{v}  \gets [~]$\;}
$PQ.enqueue((r),\wg{(r)}=0)$\;
\While{$PQ\neq \emptyset \emph{\textbf{ and }} \exists~w\neq r :  |\bestk{w}|<k$}{
$\Pi=(r,\dots,v) \gets PQ.dequeueMin()$\;
\If(\label{line:test}\tcp*[f]{Pruning Test}){$ \emph{\bf not~} \testpruning(v)$}{
\ForEach(\label{line:testcontent:start}){$u \in \nout(v) \setminus V(\Pi)$}{
$PQ.enqueue(\Pi \oplus (u),\wg{\Pi}+\omega(v,u))$\;
}
\label{line:testcontent:end}
}
\If(\tcp*[f]{Not saturated}){$|\bestk{v}|<k$}{Add $\Pi$ to $\bestk{v}$\;}
}
\caption{Algorithm \expsinglesource.}
\label{algo:basic}
\end{algorithm2e}
%%%%%%%%%%%%%testpruning PROCEDURE
\begin{algorithm2e}[t]
\small
\SetAlgorithmName{\small{Procedure}}{Procedure}
\SetAlgoLined
\KwIn{A vertex $v$.}
\KwOut{True iff all general predecessors of $v$ are saturated.}
$Q\gets \{v\}$\label{line:bfs:start}\tcp*{Add $v$ to empty FIFO queue}
$\checked \gets \{v\}$\;
\While{$Q\neq \emptyset$}{
$x\gets Q.pop()$\tcp*{Dequeue head of queue}
\If{$|\bestk{x}|<k \emph{\textbf{ and }} x \neq r$}{
\KwRet \textbf{false}\tcp*{$x$ is a general predecessor of $v$ and is not saturated}}
\ForEach{$u \in \nin(x)~:~u \not \in \checked$}{
\If{$u \in V(\bestk{x})$}{
$Q.append(u)$\label{line:bfs:predonly}\;
$\checked \gets\checked \cup \{u\}$\;
}
}
}
\KwRet \textbf{true}\;
\caption{Procedure \testpruning.}% for testing the conditions of Lemma~\ref{lem:pruning}
\label{algo:procedure:testpruning}
\end{algorithm2e}%
Our design is incremental: starting from Algorithm~\exhaustive, we first exploit the properties of the {\sc Pruning Lemma} (Lemma~\ref{lem:pruning}) to design Algorithm~\expsinglesource. In particular, note that when Algorithm \exhaustive extracts from the priority queue a path $P$ ending at a \textit{saturated} vertex $v$ (such that $|\bestk{v}|=k$); clearly, we know $P$ is not part of the top-$k$ shortest paths for $v$, but is it a prefix of a path in any solution of \ssksispshort? 
If, for some solution $\solss$, all vertices in $A_{\solss,v}$ are saturated, then by the {\sc Pruning Lemma} the answer is negative: hence, we can avoid the computation of paths with this prefix. 
By incorporating this test into Algorithm~\exhaustive we design Algorithm \expsinglesource.  
We prove its correctness, but also show that there are graphs such that its running time grows exponentially in the graph size. 
This occurs when, in the above situation, $P$ is a prefix of a path belonging to \textit{each} solution $\solss$. 
In fact, by the {\sc Path Lemma} (Lemma~\ref{lem:path}), in this case $A_{\solss,v}$ contains at least a non-saturated vertex (namely vertex $y$ in the statement of the lemma).
To overcome this limitation, we then proceed to design Algorithm~\polysinglesource which improves over Algorithm \expsinglesource by handling such pathological cases with the explicit computation of top-$k$ simple shortest paths for these non-saturated vertices, without searching for further paths having $P$ as prefix.
Thanks to this improvement, we will show that Algorithm \polysinglesource solves \ssksispshort in polynomial time w.r.t. both the graph' size and $k$.
\subsection{An Exponential-time Algorithm for \ssksispshort}
In this section, we present Algorithm \expsinglesource to solve \ssksispshort.
This algorithm, whose pseudo-code is given in Algorithm~\ref{algo:basic}, fills data structure $\bestk{v}$ with simple paths emanating from the root until a solution to the instance of \ssksispshort is achieved and works exactly as Algorithm \exhaustive except for the test of line~\ref{line:test}, which serves the purpose of reducing the search space of solutions. In particular, such test evaluates to true whenever routine $\testpruning(v)$ returns false for a given vertex $v$, terminal vertex of a path $\Pi=(r,\dots,v)$ extracted from the priority queue.
This routine (see Procedure~\ref{algo:procedure:testpruning}) checks whether all general predecessors of $v$ are saturated. In this case, by the {\sc Pruning Lemma}, the algorithm can avoid to enqueue paths that have the extracted path $\Pi$ as prefix, since these are not part of a solution.
In more detail, to determine whether all general predecessors of $v$ are saturated, Procedure~$\testpruning(v)$ performs a breadth-first-search (BFS) visit of the transpose graph that starts by enqueueing $v$ into a FIFO queue $Q$ (see line~\ref{line:bfs:start}).
Whenever a vertex $x$ is removed from $Q$ the visit proceeds by considering only the incoming neighbors of $x$ that are also its predecessors (see line~\ref{line:bfs:predonly}). 
Hence, by transitivity, it explores a set of general predecessors of $v$.
If one of the encountered vertices, say $x$, is found to be not saturated (i.e., such that $|\bestk{x}|<k$) then the routine returns false. If instead the search is concluded without finding any non-saturated general predecessor, then the routine returns true. 
Visited vertices are stored in a set \checked to avoid evaluating vertices more than once.
Algorithm~\expsinglesource stops either when the queue becomes empty or when $|\bestk{w}|=k$ for all $w\in V \setminus \{r\}$.
As in Algorithm~\exhaustive, paths in $PQ$ have a priority equal to their weight.
The correctness of Algorithm~\ref{algo:basic} is shown by the following result. 
\begin{theorem}
Algorithm~\ref{algo:basic} solves \ssksispshort. 
\label{th:basic:correct}
\end{theorem}
\begin{proof}
Observe that, by removing line~\ref{line:test}, and hence executing lines~\ref{line:testcontent:start}--\ref{line:testcontent:end} regardless of the result of the execution of procedure \testpruning, Algorithm \expsinglesource is equivalent to Algorithm \exhaustive (see Algorithm~\ref{algo:exhaustive}) which exhaustively computes a collection of top-$k$ simple shortest paths from the root vertex to all other vertices. Therefore, in this case, the claim trivially follows by Theorem~\ref{th:exhaustive:correct}.
To complete the proof, we need to show that, if the conditional statement of line~\ref{line:test} is false in some cases, and the visit does not proceed from some vertices, then the algorithm's correctness is not altered; that is, at termination, collection $\{\bestk{v_1},\bestk{v_2},\dots,\bestk{v_n}\}$ is a solution \solss to \ssksispshort, or equivalently that each $\bestk{v}$ is an element of $\collezionesol{v}$.
%%%RIMPIAZZARE W con singolo vertice u
By contradiction assume that there exists a vertex, say $u$, such that at termination, $\bestk{u}\notin\collezionesol{u}$ meaning that: either (i) $|\bestk{u}|=k'<k$ while all elements of $\collezionesol{u}$ have size $k''>k'$ with $k''\leq k$, that is collection $\bestk{u}$ is not maximal; or (ii) $|\bestk{u}|=k$ but there exists a path $P'$ in the graph that has been added to $\bestk{u}$ whereas it should not have been, that is there exists at least a path $P$  in the graph such that $\wg{P}<\wg{P'}$.
In both cases, it follows that there exists at least a path in the graph that should be in \bestk{u} but is not added to \bestk{u} during the execution of the algorithm.
%
% Now, call $M$ the set of all such paths and l
Among these paths, call $P_{ru}$ the one that contains a vertex $v \in P_{ru}$ such that $v$ is the first vertex for which the test in line~\ref{line:test} is false, that is, the first vertex during the execution of the algorithm such that $\testpruning(v)$ returns true.
If we denote by $X$ the set of vertices that are visited by routine $\testpruning(v)$ executed in line~\ref{line:test}, then, for any $x\in X$ we have that: (i) $x$ is saturated, i.e. $|\bestk{x}|=k$; (ii) $\bestk{x}\in \collezionesol{x}$ by Theorem~\ref{th:exhaustive:correct}. Thus, set $X \equiv A_{\solss,v}$ for some solution \solss. 
Hence, according to Lemma~\ref{lem:pruning}, there must exist a solution $\solss'=\{\solssvertex{v_1}',\solssvertex{v_2}',\ldots,\solssvertex{v_n}'\}$ such that $P_{rv}$ is not a prefix of any path in $\underset{w \in V}\bigcup\solssvertex{w}'$, which contradicts the hypothesis of $P_{rv}$ being a prefix of $P_{ru}$ and $P_{ru} \subseteq \solssvertex{u}$ for any $\solssvertex{u}\in \collezionesol{u}$.
\end{proof}
Clearly, intuitively, the pruning test of line~\ref{line:test} should allow one to reduce the number of paths that Algorithm \expsinglesource enqueues in (dequeues from, respectively) the priority queue to determine a solution to \ssksispshort, compared to Algorithm \exhaustive.
However, the following result shows that the introduction of the pruning test does not suffice to guarantee a polynomial time complexity for Algorithm~\expsinglesource:
\begin{theorem}
There exist an $n$-vertex graph and an integer $k>0$ that force algorithm \expsinglesource to run in $\Omega(2^n)$ time to achieve a solution to \ssksispshort.
\label{th:basic:runningtime}
\end{theorem}
\begin{proof}
Consider the graph of Figure~\ref{fig:basic:exponential} and analyze the case of $k=3$ with $r$ as root vertex. 
Observe that all simple shortest paths from $r$ to vertex $x_1$ can be classified into two categories: in former we have (two) simple paths that do not share any vertex with the internal vertices of the red dashed path, namely $(r,x_1)$, $(r,x_2,c_1,x_1)$; in the latter we have all simple paths formed by concatenating all simple paths from $r$ to vertex $v$ with the red path from $v$ to $x_1$.
If we call $d$ the number of vertices of type $c_i$ in the given graph, then we have $2^{d+1}$ of simple paths from $r$ to $x_1$ of the latter category. 
Now, assume the dashed red path has weight $2d$. Then, Algorithm \expsinglesource finds the two simple paths of the first category after at most three dequeue and at most nine dequeue operations, respectively, depending on the order in which neighbors are considered.
Similarly, after at most nine dequeue operations, the paths terminating at $x_3$ and $x_4$ are dequeued, and the four paths $(r,x_1,c_1,x_3,c_2)$, $(r,x_1,c_1,x_4,c_2)$, $(r,x_2,c_1,x_3,c_2)$, $(r,x_2,c_1,x_4,c_2)$ terminating at $c_2$ are then enqueued. Observe that such enqueue operations follow from the test of line~\ref{line:test} evaluating to true for both $x_3$ and $x_4$: this occurs since $\bestk{x_3}$ and $\bestk{x_4}$ have size $k=3$ and both calls to routine $\testpruning(x_3)$ and $\testpruning(x_4)$ return false  -- due to $x_1$ and $x_2$ being not saturated.
This behavior continues unchanged for all vertices in all simple paths from $r$ to $v$ that do not share vertices with the red path, with routine $\testpruning$ always returning false since $x_1$ is a predecessor of all such vertices and is not saturated. 
Hence, $2^{d+1}=\bigO(2^n)$ paths terminating at $v$ are enqueued and dequeued before any of the $2^{d+1}$ simple paths from $r$ to $x_1$ of the second category can be enqueued/dequeued, and therefore before the algorithm can successfully terminate.
\end{proof}
\subsection{Solving \ssksispshort in Polynomial Time}
In this section, we present Algorithm \polysinglesource that solves \ssksispshort in polynomial time.
The algorithm, whose pseudo-code is given in Algorithm~\ref{algo:polynomial}, is built upon Algorithm \expsinglesource and uses the characterization of solutions to \ssksispshort of Section~\ref{sec:char} to limit the maximum number of paths that are enqueued and hence dequeued from the priority queue, for each vertex other than the root.
\begin{algorithm2e}[t]
\small
\SetAlgoLined
\KwIn{Graph $G=(V,E,\omega)$, root $r \in V$, $k\in \mathbb{N}$.}
\KwOut{Top-$k$ simple shortest paths $\bestk{v}$ from $r$ to each $v \in V \setminus\{r\}$.}
$PQ\gets \emptyset$\tcp*{Empty priority queue} 
\ForEach(\tcp*[f]{Empty lists}){$v\in V$}{$\bestk{v}  \gets [~]$\:}
$PQ.enqueue((r),\wg{(r)}=0)$\;
\While{$PQ\neq \emptyset \emph{~\bf and~} \exists~w\neq r :  |\bestk{w}|<k$\label{line:loop} }{
$\Pi=(r,\dots,v) \gets PQ.dequeueMin()$\label{line:dequeue}\;
\If(\label{line:nonsat}){$|\bestk{v}|<k$}{
Add $\Pi$ to $\bestk{v}$\label{line:insert}\;
\ForEach(\label{line:forloop}){$u \in \nout(v)\setminus V(\Pi)$}{
\If(\label{line:iftest}){$u$ not super-saturated $\emph{\bf and~}~ \Pi \oplus (u) \not \in PQ$}{
$PQ.enqueue(\Pi \oplus (u),\wg{\Pi}+\omega(v,u))$\label{line:enqueue_normal}\;
}
}
}
\ElseIf(\label{line:extra}){$v$ not super-saturated\label{line:notssat}}{
\ForEach{$w$ general predecessor of $v$}{
Compute $\solssvertex{w}\supseteq \bestk{w}$\label{line:notssatcalc}\;
\ForEach{$p \in \solssvertex{w}\setminus \bestk{w}~:~p \not \in PQ$\label{line:forasv}}{
$PQ.enqueue(p,\wg{p})$\label{line:enqueue_asv}\;}
Mark $w$ \textit{super-saturated}\label{line:supersat}\;
}
}
}
\caption{Algorithm \polysinglesource.}
\label{algo:polynomial}
\end{algorithm2e}
To obtain such a bound, the main modification consists in replacing procedure \testpruning, which searches for a non-saturated general predecessor to decide whether a visit must be pruned at a given path $\Pi=(r,\dots,v)\notin \solssvertex{v}$ or not, with an explicit computation of top-$k$ simple shortest paths for non-saturated general predecessors of $v$. 
Consider again the graph in Fig.~\ref{fig:basic:exponential} and assume $k=3$. Define a vertex $x$ to be \emph{super-saturated} whenever the algorithm has computed a collection of top-$k$ simple shortest paths from the root to each general predecessor of $x$, including $x$ itself.
When Algorithm \polysinglesource extracts the fourth among the paths terminating at $c_2$ -- namely $(r,x_1,c_1,x_3,c_2)$, $(r,x_1,c_1,x_4,c_2)$, $(r,x_2,c_1,x_3,c_2)$, and $(r,x_2,c_1,x_4,c_2)$ -- then $|T_{c_2}|=k$ and $c_2$ is not supersaturated. Thus, in line~\ref{line:notssatcalc}, the algorithm immediately computes top-$k$ simple shortest paths $\solssvertex{x_1}$ and $\solssvertex{x_2}$ for vertices $x_1$ and $x_2$, predecessors of $c_2$, in such a way that $\solssvertex{x_1}$ and $\solssvertex{x_2}$ include the paths already stored in $T_{x_1}$ and $T_{x_2}$. This avoids the computation of an exponential number of paths before completing the computation of a solution to \ssksispshort, as instead happens if \expsinglesource is used. 
In more detail, the algorithm works as follows.
The initialization phase is as in Algorithm \expsinglesource: we set up $\bestk{v}$ to be an empty list, for each $v\in V$, and priority queue $PQ$ to contain only path $(r)$, for the root vertex $r\in V$, with priority equal to $0$. 
Then, as long as $PQ$ is not empty or $|\bestk{w}|<k$ for some $w\in V\setminus\{r\}$, a path $\Pi$ terminating at some vertex $v$ is dequeued from $PQ$ and an \textit{iteration} of the algorithm is executed.
In a generic iteration, if the conditional statement of line~\ref{line:test} evaluates to true (i.e. no pruning is possible) then the execution proceeds as in Algorithm \expsinglesource, with the difference that simple paths $\Pi\oplus (u)$ are enqueued into $PQ$ only for the vertices $u$ in $\nout(v)$ that are not \textit{super-saturated}.
In particular, initially all vertices are not super-saturated.
Then, when the test of line~\ref{line:test} evaluates to false, that is when we extract a path $\Pi=(r,\dots,v)$ and $|\bestk{v}| = k$ we distinguish two cases: either $v$ is super-saturated or not. 
In the former case, no further enqueue operation is performed, for the same reasons that stop the visit of Algorithm \expsinglesource when \testpruning returns true.
In the latter, instead, we compute explicitly a set of general predecessors of $v$ and a collection $\solssvertex{w}$ that contains paths in $\bestk{w}$, for each general predecessor $w$ of $v$: on the one hand, this is easy to achieve for saturated vertices, for which we will show $\bestk{w}=\solssvertex{w}$ for some $\solssvertex{w}\in \collezionesol{w}$; on the other hand, it requires to compute a collection of top-$k$ simple shortest paths from $r$ to $w$, i.e., a solution to \spksispshort for pair $(r,w)$, for each non-saturated general predecessor $w$ of $v$. 
Such a solution can be obtained by any algorithm that solves \spksispshort for pair $(r,w)$ in polynomial-time (e.g., Yen's).
Once this is done, we enqueue any path $p \in \solssvertex{w}\setminus \bestk{w}$ into $PQ$ and mark every general predecessor of $v$ to be super-saturated.
Observe that, whenever we need to add a path $p$ to $PQ$ during the algorithm, we verify that $p \not \in PQ$ before performing the enqueue operation, to avoid inserting twice a same path in $PQ$ and to achieve a correct result.
The correctness of Algorithm~\ref{algo:polynomial} is given by the following result. 
\begin{theorem}
\label{thm:polynomial:correctness}
Algorithm~\ref{algo:polynomial} solves \ssksispshort.
\end{theorem}
\newcommand{\tk}[2]{\ensuremath{\bestk{#1}^{#2}}}
%%%%%%%%%%%%%%
\begin{proof}    
In what follows, we use \tk{u}{t} to represent the content of data structure \bestk{u} at the $t$-th iteration of the main loop of line~\ref{line:loop}. Similarly,  we denote by $PQ^t$ the priority queue $PQ$ after the $t$-th execution of such a loop.
We assume $t=0$ if no loop iteration has been performed and $t=1$ when one iteration has been performed.
If $A$ is a collection of paths, we call $L(A)$ the \emph{profile} of $A$, that is the ordered list of the weights of the paths in $A$. Hence, a pair of profiles can be naturally compared by using the lexicographical ordering, i.e. for any two profiles $L(A')$ and $L(A'')$ of two collections of paths $A'$ and $A''$ we write comparison expressions such as $L(A')\geq L(A'')$, $L(A')<L(A'')$ or $L(A')=L(A'')$.
%%
%%%%%%%%%%%%%%%%%%%%%%%%%
First, observe that $L(\solssvertex{u})=L(\solssvertex{u}')$ even if $\solssvertex{u}$ and $\solssvertex{u}'$ are different solutions to \ssksispshort for a given vertex $u$, i.e. $\solssvertex{u} \in \collezionesol{u}$ and $\solssvertex{u}' \in \collezionesol{u}$ with $\solssvertex{u}\neq \solssvertex{u}'$. 
Moreover, if $L(\tk{u}{t})\leq L(\solssvertex{u})$ then all paths in $\tk{u}{t}$ must belong to a solution in $\collezionesol{u}$ for vertex $u$. Finally, if $L(\tk{u}{t})= L(\solssvertex{u})$ then $\tk{u}{t}$ belongs to a solution $\solssvertex{u}$ for $u$.
\par
The proof is by induction on the number $t$ of times that the main loop of line~\ref{line:loop} is performed. We use  $P_{ru}$ to denote a generic path from a vertex $r$ to a vertex $u$ in the graph.
We prove that, at each time $t$ and for each $w \in V$, the following two properties hold:
\begin{enumerate}
    \item $L(\tk{w}{t})\leq L(\solssvertex{w})$;
    \item if there exists a path $P_{rw}$ such that $P_{rw} \not \in \tk{w}{t}$ and $L(\tk{w}{t}\cup \{P_{rw}\})\leq L(\solssvertex{w})$ then there exists a prefix $P'$ of $P_{rw}$ such that $P'\in PQ^t$.
\end{enumerate}
On the one hand, the first property ensures that paths in $\tk{w}{t}$ computed until time $t$ are correct, i.e. paths in $\tk{w}{t}$ belong to a solution to \ssksispshort for each vertex $w\in V$, and that such paths are found in the due non-decreasing order of weight. 
The second property, on the other hand, guarantees that any next path that has to be added to $\tk{w}{t}$ to complete a solution in $\collezionesol{w}$ has a prefix in $PQ^t$ for each vertex $w$. Note that by the definition of prefix, any path is a prefix of itself.
If we focus on time $t=0$, we have that $L(\tk{w}{0})=[~]$ for each vertex $w\in V$ while $PQ^0=\{(r)\}$. Hence, both properties 1) and 2) hold since path $(r)$ is prefix of any path in a solution of \ssksispshort.
\par
Now we assume that properties 1) and 2) are true at time $t$ and show that they remain true after the execution of the loop for the $(t+1)$-th time.
We first consider the case of the condition controlling the main loop being true, then $PQ^t$ is not empty.
Let $P_{rv}$ the path dequeued in line~\ref{line:dequeue}. Notice that, like $P_{rv}$, any path in $PQ$ can be inserted either in line~\ref{line:enqueue_normal} or in line~\ref{line:enqueue_asv}. We call \emph{normal} (\emph{exceptional}, respectively) the first (second, respectively) kind of insertion.
\par\underline{We first prove that property 1) holds at time $t+1$.}
We distinguish two cases. If $P_{rv}$ was inserted in $PQ$ as a consequence of an exceptional insertion, then we have $L(\tk{v}{t+1})=L(\tk{v}{t}\cup \{P_{rv}\})\leq L(\solssvertex{v})$, as exceptional insertions are dictated by an execution of Yen's (or any algorithm for \spksispshort) algorithm which returns a solution $\solssvertex{v}$ for any vertex $v$. Moreover, since $L(\tk{w}{t+1})=L(\tk{w}{t})$ for any other vertex $w\neq v$, then property 1) holds.
Consider, instead, that $P_{rv}$ was inserted in $PQ$ as a consequence of a normal insertion. If $|L(\tk{v}{t})|=|L(\solssvertex{v})|$ it follows that property 1) clearly remains true as $P_{rv}$ is not added to $\tk{v}{t}$. 
If, instead, $|L(\tk{v}{t})|<|L(\solssvertex{v})|$, we suppose, by contradiction, that $L(\tk{v}{t}\cup \{P_{rv}\}) \not \leq L(\solssvertex{v})$. Since $|L(\tk{v}{t})|<|L(\solssvertex{v})|$, there must exist a path $P_{rv}'$ such that $L(\tk{v}{t}\cup \{P_{rv}'\})  \leq L(\solssvertex{v})$ and we can distinguish  two cases, since $\wg{P_{rv}}\neq \wg{P_{rv}'}$: either (a) $\wg{P_{rv}}>\wg{P_{rv}'}$; or (b)~$\wg{P_{rv}}<\wg{P_{rv}'}$.
% \begin{enumerate}
%     \item[a)]~$\wg{P_{rv}}>\wg{P_{rv}'}$;
%     \item[b)]~$\wg{P_{rv}}<\wg{P_{rv}'}$.
% \end{enumerate}
\par For case (a): since $L(\tk{v}{t}\cup \{P_{rv}'\})\leq L(\solssvertex{v})$, then, by property 2),  there exists a prefix $P'$ of $P_{rv}'$ such that $P'\in PQ^t$.
Since $P'$ is a prefix of $P_{rv}'$ we have  $\wg{P'}\leq \wg{P_{rv}'}$. Moreover $\wg{P_{rv}'}<\wg{P_{rv}}$, which implies that $P'$ should have been extracted from $PQ$ in place of $P_{rv}$, since we extract in order of weight.
\par
For case (b): If $\wg{P_{rv}}<\wg{P_{rv}'}$ then, by property 1) at time $t$, $P_{rv}$ is already in $\tk{v}{t}$ and then the algorithm has inserted $P_{rv}$ in $PQ$ twice.
This leads to a contradiction. 
In fact, first of all, there cannot be two insertions of $P_{rv}$ that are both normal, since this would imply that prefix $P''$ of $P_{rv}$, obtained by removing $v$ from $P_{rv}$ would have been inserted twice in $PQ$ in line~\ref{line:insert}, which in turn implies $L(\tk{v}{t})\not \leq L(\solssvertex{v})$.
Moreover, if first insertion of $P_{rv}$ is exceptional and the second one is normal, then $v$ is marked \textit{super-saturated} by the algorithm and no path terminating at $v$ is inserted again in $PQ$ (see lines~\ref{line:iftest} and~\ref{line:notssat}).
Finally, if the first insertion of $P_{rv}$ is normal and the second is exceptional, we obtain a contradiction since an exceptional insertion concerns only paths in some solution $\solssvertex{v}$ that are not already in \tk{v}{} (see line~\ref{line:forasv}).
\par
\underline{We now prove that property 2) holds at time $t+1$.}
Assume first that the condition of line~\ref{line:nonsat} is true. Hence, path $P_{rv}$ is removed from $PQ$ but, for each non super-saturated vertex $u$ in $\nout(v)\setminus V(P_{rv})$, path $P_{ru}=P_{rv} \oplus (u)$, is inserted in $PQ$. 
Then, if $u$ is not super-saturated, we have that $P_{ru}$ is now in $PQ^{t+1}$ and it is a prefix of any path $P_{rw}$ from the root vertex to a vertex $w$ that had $P_{rv}$ as prefix. 
Vice versa, if $u$ is super-saturated, the algorithm correctly does not enqueue $P_{ru}$ in $PQ^{t}$. Indeed, $P_{ru}\not \in \solssvertex{u}$ and if there were a path $P_{rw}$, as in property 2), having $P_{ru}$ as prefix, then by Lemma~\ref{lem:path} there would exist a vertex $y$ in $P_{rw}$ such that a path $P_{ry} = P_{ru}\oplus P_{uy}$ is in $\solssvertex{y}$, for some solution $\solssvertex{y}$. 
By the same lemma, $y$ is a general predecessor of $u$, hence path $P_{ry}$ has been determined and added to $PQ$ via an exceptional insertion when $u$ has become super-saturated. Path $P_{ry}$ is a prefix of $P_{rw}$ and, since $\wg{P_{ry}}> \wg{P_{rv}}$ and $PQ$ is a priority queue, $P_{ry}$ is still in $PQ^{t+1}$.
% }
\par
Now consider the case when the condition of line~\ref{line:nonsat} is false.
If $v$ is super-saturated, then simply $P_{rv}$ is removed from $PQ$ and no path is inserted. However, if $P_{rv}$ was a prefix of a path $P_{rw}$, as above, any prefix of $P_{rw}$ has been computed and inserted as exceptional in $PQ$ when $v$ has become super-saturated.
If $v$ is not super-saturated and $P_{rv}$ is a prefix of a path $P_{rw}$, again by Lemma~\ref{lem:path} there exists a vertex $y$ in $P_{rw}$ such that $P_{ry}=P_{rv}\oplus P_{vy}$ is in $\solssvertex{y}$, for some solution $\solssvertex{y}$. As $y$ is in the set of the general predecessors of $v$, then path $P_{ry}$ is inserted in $PQ^{t+1}$ in this iteration of the loop and this suffices to show that property 2) holds.

To complete the proof we now consider the case when the condition controlling the main loop (see line~\ref{line:loop}) is false at time $t_e$. If $\forall~w\neq r :  |\tk{w}{t_e}|=k$ then, by property 1), the algorithm has computed a correct solution for \ssksispshort. If instead $PQ^{t_e}$ is empty, since there is no path in $PQ^{t_e}$ then the conclusion of property 2) is false which implies that the premise is also false. Hence, for each path $P_{rw}$ we have either $P_{rw}\in \tk{w}{t_e}$ or $L(\tk{w}{t_e}\cup \{P_{rw}\}) \not \leq L(\solssvertex{w})$. In both cases it follows that $L(\tk{w}{t_e}) = L(\solssvertex{w})$ and therefore that the algorithm has computed a correct solution for \ssksispshort.
\end{proof}
Now, if we call $\textsc{sp}(n,m)$ the time complexity of a shortest-path algorithm in an $n$-vertex $m$-edge graph, the time complexity of Algorithm~\ref{algo:polynomial} is as follows:
\begin{theorem}
Algorithm~\polysinglesource runs in $\bigO(k(m\log km+n^2 \textsc{sp}(n,m)))$ time.
\label{thm:polynomial:runningtime}
\end{theorem}
\begin{proof}
Observe that the number of times the loop in line~\ref{line:loop} of algorithm~\ref{algo:polynomial} is executed is upper bounded by the number of insertions in the queue $PQ$, which is given by the sum of the number of normal insertions and the number of exceptional insertions.
The former is $\bigO(km)$ since, for each vertex $v\in V$, we perform at most $k\delta_v$ insertions (see line~\ref{line:enqueue_normal}), where $\delta_v=|\nout(v)|$ is the outdegree of the vertex. By summing up for all vertices, we have $\sum_{v\in V} k\delta_v=k\sum_{v\in V} \delta_v=2km=\bigO(km)$.
The latter, instead, is $\bigO(nk)$ since: (i) in the worst case, the number of general predecessors of a vertex is at most $n$; (ii) we execute 
an algorithm for \spksispshort and enqueue at most $k$ paths per general predecessor.
Now, for each insertion, the algorithm enqueues a path in $PQ$ which is a priority queue whose size is at most the total number of insertions, bounded by $\bigO(nk+mk)$. Thus, each insertion requires $\bigO(\log (k(n+m)))$ operations and hence the running time due to insertions is $\bigO(k(m+n)\log (k(m+n)))$. Similarly, the  running time for corresponding dequeue operations is  $\bigO(k(m+n)\log (k(m+n)))$. 
Moreover, for each of the at most $n-1$ vertices for which we perform at most $k$ exceptional insertions, we need to run an algorithm for solving \spksispshort. For this task, if we select Yen's algorithm, the best algorithm in terms of time complexity for this purpose, which runs in $\bigO(n\cdot kn \cdot \textsc{sp}(n,m))=\bigO(kn^2 \textsc{sp}(n,m))$ time~\cite{ZoobiCN23}, we obtain a total running time for algorithm \polysinglesource of $\bigO(k(m+n)\log (k(m+n))+kn^2 \textsc{sp}(n,m))=\bigO(km\log km+kn^2\textsc{sp}(n,m))=\bigO(k(m\log km+n^2 \textsc{sp}(n,m)))$.
\end{proof}
Observe that, for unweighted graphs, $\textsc{sp}(n,m)$ is the running time of the breadth-first-search algorithm, which is $\bigO(m+n)$. Vice versa, for weighted graphs, $\textsc{sp}(n,m)$ is upper bounded by the running time of Dijkstra's algorithm, i.e. $\bigO(m+n\log n)$.
Thus, the following corollary can be derived:
\begin{corollary}
The running time of Algorithm~\ref{algo:polynomial} is asymptotically upper bounded by the running time of $n-1$ executions of Yen's algorithm.
\label{cor:polynomial:runningtime_upper}
\end{corollary}
\begin{proof}
Observe that Algorithm \polysinglesource runs for~$T(n,m,k)=\bigO(k(m\log km+n^2 \textsc{sp}(n,m)))$ time.
If we expand the product, we have that $T(n,m,k)=\bigO(k(m\log km+n^2 \textsc{sp}(n,m)))=\bigO(km\log km+kn^2\textsc{sp}(n,m))$. Since $\textsc{sp}(n,m)$ is $\bigO(m+n\log n)$ for weighted graphs, we obtain that $T(n,m,k)=\bigO(km\log km+kn^2(m+n \log n))$. Moreover, notice that, in the worst case, $k$ can be exponential in $n$, therefore $\log km=\bigO(n+\log m)$ and hence $km\log km = \bigO(km(n+\log m))$. This implies that, in any graph such that $m =\Omega(n)$, $\bigO(km(n+\log m))$ is upper bounded by $\bigO(kn^2(m+n \log n))$ and therefore by the time taken by $\Theta(n)$ executions of Yen's algorithm. A similar result can be obtained if one considers unweighted graphs, by replacing $\textsc{sp}(n,m)$ with $\bigO(m+n)$.
\end{proof}
Note that Corollary~\ref{cor:polynomial:runningtime_upper} holds for any polynomial-time algorithm for \spksispshort that matches Yen's time complexity (e.g. \pnc~\cite{ZoobiCN23} or that in~\cite{KM16}).
Note also that a lower level, more detailed description of Algorithm \polysinglesource is provided in Algorithm~\ref{algo:polynomial:implemented}. In detail, set $A_{\solss,v}$ is computed through an iterative computation of the transitive closure of the relation predecessors (see Definition~\ref{def:general_predecessor}) that uses a membership set \checked to terminate and not to visit a vertex more than once (note that a vertex can be a predecessor of many vertices in a solution). 
In such a description, we use notation $\spkalgo(G,r,x,k)$ to denote a generic call to an algorithm that solves \spksispshort that returns a collection of top-$k$ simple shortest paths for a pair $(r,x)$ of vertices in graph $G$. 
Super-saturated vertices are stored in a set named \supersat.
\begin{algorithm2e}[ht]
\small
\SetAlgoLined
\small
\SetAlgoLined
\KwIn{Graph $G=(V,E,\omega)$, root $r \in V$, $k \in \mathbb{N}$.}
\KwOut{Top-$k$ simple shortest paths $\bestk{v}$ from $r$ to each $v \in V \setminus\{r\}$.}
$PQ\gets \emptyset$\tcp*{Empty priority queue} 
\ForEach(\tcp*[f]{Empty lists}){$v\in V$}{$\bestk{v}  \gets [~]$\;}
$PQ.enqueue((r),\wg{(r)}=0)$\;
$\supersat \gets \{r\}$\tcp*{Super-saturated vertices}
\While{$PQ\neq \emptyset \emph{~\bf and~} \exists~w\neq r :  |\bestk{w}|<k$}{
$\Pi=(r,\dots,v) \gets PQ.dequeueMin()$\tcp*{Min-weight path} 
\If{$|\bestk{v}|<k$}{
Add $\Pi$ to $\bestk{v}$\;
\ForEach{$u \in \nout(v)\setminus V(\Pi)$}{
\If{$u\not\in \supersat$ {\normalfont \textbf{and}} $\Pi \oplus (u) \not \in PQ$}{
$PQ.enqueue(\Pi \oplus (u),\wg{\Pi}+\omega(v,u))$\;
}
}
}
\ElseIf{$v\not\in \supersat$}{
$A_{\solss,v}\gets \emptyset$\;
$Q\gets \{v\}$\tcp*{Add $v$ to empty FIFO queue}
$\checked \gets \{v\}$\;
\While{$Q\neq \emptyset$}{
$x\gets Q.pop()$\tcp*{Dequeue head of queue}
\If{$x\not\in \supersat$}{
\lIf{$|\bestk{x}|<k$}{$\solssvertex{x} \gets V(\spkalgo(G,r,x,k))$\label{line:yen}}
\lElse{$\solssvertex{x} \gets \bestk{x}$}
$A_{\solss,v}\gets A_{\solss,v} \cup V(\solssvertex{x})$\;
\ForEach{$p \in \solssvertex{x}\setminus \bestk{x}~:~p \not \in PQ$}{
$PQ.enqueue(p,\wg{p})$\;
}
\ForEach{$w \in V(\solssvertex{x})$}{
\If{$w\not\in \checked$}
{
\If{$w\not\in \supersat$}{
$\checked \gets \checked\cup \{w\}$\;
$Q.append(w)$\;
}
}
$\supersat\gets \supersat\cup\{w\}$\;
}
$\supersat\gets \supersat\cup\{x\}$\;

}
}
}
}
\caption{Detailed description of Algorithm~\ref{algo:polynomial}.}
\label{algo:polynomial:implemented}
\end{algorithm2e}

\section{Experimental Evaluation}\label{sec:experiments}
In this section, we present the results of an extensive experimental evaluation aimed at assessing the average performance of Algorithm~\polysinglesource.
\begin{table*}[t]
\centering
\renewcommand{\arraystretch}{0.9} % this reduces the vertical spacing between rows
\subfloat[\normalsize  Input graphs used in trials with $k\in\{2,4,8,16\}$.]{
\resizebox{\textwidth}{!}{ 
\begin{tabular}{c|c|c|rr|cc|rrr|c}
\textbf{Dataset} & \textbf{graph} & \textbf{Type} & $|V|$ & $|E|$ & \textbf{D} & \textbf{W} & \avgdeg  & \maxdeg & $\Delta$ & \textbf{S} \\
\midrule
\linux & \linuxss & \textsc{blogging} & 913 & 4162 & \true{2pt} & \false{2pt} & 4.56  & 61 & 6 & \false{2pt} \\
\citat & \citatss & \textsc{collabor.} & 4158 & 13422 & \false{2pt} & \false{2pt} & 6.46  & 81 & 17 & \false{2pt} \\
\slashd & \slashdss & \textsc{friendship} & 2196 & 22773 & \true{2pt} & \false{2pt} & 10.37 & 1991 & 4 & \false{2pt} \\
\caida & \caidass & \textsc{ethernet} & 32000 & 40204 & \false{2pt} & \true{2pt} & 2.51  & 203 & 61 & \false{2pt} \\
\peer & \peerss & \textsc{peer2peer} & 14149 & 50916 & \true{2pt} & \false{2pt} & 3.60 & 32 & 9 & \false{2pt} \\
\lux & \luxss & \textsc{road network} & 30647 & 75546 & \true{2pt} & \true{2pt} & 2.47 & 9 & 490 & \false{2pt} \\
% \wikiv & \wikivss & \textsc{voting} & 7066 & 100736 & \false{2pt} & \false{2pt} & 28.51 & 1065 & 7 & \false{2pt} \\
\amazon & \amazonss & \textsc{rating} & 20168 & 104513 & \false{2pt} & \false{2pt} & 10.36 & 169 & 31 & \false{2pt}\\

\emaild & \emaildss & \textsc{email} & 34203 & 151132 & \true{2pt} & \false{2pt} & 4.42  & 715 & 13 & \false{2pt} \\

\bright & \brightss & \textsc{location} & 56739 & 212945 & \false{2pt} & \false{2pt} & 7.51 & 1134 & 30 & \false{2pt} \\

\youtub & \youtubss & \textsc{media} & 28372 & 249511 & \false{2pt} & \false{2pt} & 17.59  & 8920 & 13 & \false{2pt} \\
\barab & \barabss & \textsc{random power law} & 71920 & 299480 & \true{2pt} & \false{2pt} & 4.16  & 571 & 17 & \true{2pt} \\
\moc & \mocss & \textsc{rand. spatial} & 18848 & 318248 & \true{2pt} & \true{2pt} & 16.88  & 160 & 1168 & \true{2pt} \\
\arxiv & \arxivss & \textsc{citation} & 34401 & 420784 & \true{2pt} & \false{2pt} & 24.46 & 15 & 846 & \false{2pt} \\
\cisco & \ciscoss & \textsc{distributed system} & 30181 & 424232 & \true{2pt} & \false{2pt} & 14.06 & 18589 & 12 & \false{2pt} \\

\faceb & \facebss & \textsc{social network} & 37165 & 494168 & \false{2pt} & \false{2pt} & 26.59 & 950 & 17 & \false{2pt} \\
\epi & \episs & \textsc{reviews} & 41441 & 693507 & \true{2pt} & \false{2pt} & 16.73  & 1820 & 17 & \false{2pt} \\
% \wkt & \wktss & \textsc{chatnet} & 25411 & 902999 & \true{2pt} & \false{2pt} & 35.54 & 74238 & 11 & \false{2pt} \\
\biolog & \biologss & \textsc{biological} & 4858 & 982944 & \false{2pt} & \true{2pt} & 404.67 &  2242 & 94555 & \false{2pt} \\
\direrdos & \direrdosss & \textsc{random uniform} & 10000 & 999809 & \true{2pt} & \false{2pt} & 99.98  & 138 & 3 & \true{2pt} \\
\end{tabular}
}
\label{table:input:subset1}
}\\
% \begin{table*}[t]
% \centering
% \renewcommand{\arraystretch}{0.8} % this reduces the vertical spacing between rows
\subfloat[\normalsize Input graphs used in trials with $k \in \{2^i\}_{i=1,2,\dots,10}$.]{
\resizebox{\textwidth}{!}{ 
\begin{tabular}{c|c|c|rr|cc|rrr|c}
\textbf{Dataset} & \textbf{graph} & \textbf{Type} & $|V|$ & $|E|$ & \textbf{D} & \textbf{W} & \avgdeg  & \maxdeg & $\Delta$ & \textbf{S} \\
\midrule
\ita & \itass & \textsc{train network} & 1329 & 3862 & \true{2pt} & \true{2pt} & 2.91 &  13 & 890 & \false{2pt} \\
\scf & \scfss & \textsc{random scale free} & 1050 & 4112 & \true{2pt} & \false{2pt} & 3.92 & 201 &  14 &\true{2pt} \\
\bitcoin & \bitcoinss & \textsc{financial transactions} & 5875 & 21489 & \false{2pt} & \true{2pt} & 7.32 &  795 & 34 & \false{2pt} \\
\oregon & \oregonss & \textsc{autonomous system} & 10670 & 22002 & \false{2pt} & \false{2pt} & 4.12 & 2312 & 10 & \false{2pt} \\
% \openfli & \openfliss & \textsc{flight network} & 2868 & 30404 & \true{2pt} & \false{2pt} & 10.60  & 237 & 14 &  \false{2pt} \\
\google & \googless & \textsc{web index} & 12354 & 164046 & \true{2pt} & \false{2pt} & 13.28 & 120 & 6 & \false{2pt} \\
\end{tabular}
}
\label{table:input:subset2}
}
% \caption{Overview of graphs used for tests with $k \in \{2^i\}_{i=1,2,\dots,10}$: columns are as in Table~\ref{table:input}}
% \label{table:input:scale}
% \end{table*}

\caption{Overview of used input graphs: columns 1st to 3rd provide dataset name, graph acronym, type of network; columns 4th and 5th report number of vertices and arcs while columns 6th (\textbf{D}) and 7th (\textbf{W}) indicate whether the graph is directed and/or weighted (\true{2pt} = true, \false{2pt} = false);
columns 8th (\avgdeg), 9th (\maxdeg) and 10th ($\Delta$) contain, respectively, the average and maximum (outgoing) degree and the diameter of the graph; the  last column (\textbf{S}) shows whether the graph is synthetic (\true{2pt}) or real (\false{2pt}). Inputs are sorted by $|E|$, non-decreasing.
}
\label{table:input}
\end{table*}
In fact, although its worst-case time complexity matches that of the best-known polynomial-time solution to \ssksispshort, obtained by running a polynomial-time \spksispshort algorithm once for each vertex pair $(r, v)$, our method can often achieve significantly better performance in practice, mainly thanks to the effectiveness of the strategy that exploits shared path prefixes to determine different top-$k$ collections from the root to other vertices.
Thus, our main goal is to evaluate the average-case effectiveness of this strategy and determine whether Algorithm~\polysinglesource outperforms the naive baseline, which applies a polynomial-time \spksispshort algorithm repeatedly. This comparison is critical to establish whether \polysinglesource constitutes an improvement over the state-of-the-art for \ssksispshort.
To this end, we implemented \polysinglesource and compared it with a generalization of Yen's algorithm~\cite{Yen71}, which solves \ssksispshort by running Yen’s method $n-1$ times, once for each pair $(r, v)$, with $r \in V$ and $v \in V \setminus {r}$. We refer to this baseline as Algorithm~\ssyen.
For completeness, we also include a second baseline that also solves \ssksispshort by executing $n-1$ times algorithm \pnc~\cite{ZoobiCN23}, to handle \spksispshort for the same pairs. This baseline is selected since algorithm \pnc has been shown empirically to outperform Yen's, for \spksispshort,  in several real-world scenarios, despite sharing the same worst-case time complexity.
We refer to this variant as Algorithm~\sspnc. 
For improved performance, our implementation of \pnc integrates the efficient heuristic introduced in~\cite{FengCLJ23}, which reduces the search effort by computing an upper bound on the weight of the $k$-th heaviest shortest path.

We do not consider approaches for \spksispshort whose performance is generally dominated by or comparable to that of \pnc~\cite{Feng14,Gao10}, and methods with high space overhead (due to storing many shortest-path trees to accelerate the computation)~\cite{KM16}.   Moreover, we exclude from our evaluation the algorithm of~\cite{katoh1982efficient} specific for \spksispshort in undirected graphs, since empirical studies have found its implementation to be often slower than that of Yen's~\cite{pascoal2006implementations}.
For fairness, we implemented and tested two variants of Algorithm~\polysinglesource: one using Yen's algorithm to compute general predecessors (see line~\ref{line:notssatcalc} of Algorithm~\ref{algo:polynomial} or line~\ref{line:yen} of Algorithm~\ref{algo:polynomial:implemented}), and another using \pnc. We refer to these as Algorithms~\shortpolyyen and~\shortpolypnc, respectively.

All algorithms have been implemented with support for both undirected and unweighted graphs. 
This is achieved by removing edge orientations (i.e. the distinction between $\nin$ and $\nout$) or by assuming unit weights and replacing Dijkstra-like traversals with BFS-like ones, as in previous works on shortest paths~\cite{Eppstein98,AkibaHNIY15,DD23,DEmidioFFLP19}.
For completeness, we implemented and tested algorithm \expsinglesource as well. 
However, preliminary experimentation reveals that its running time is not infrequently huge even for small graphs and $k$, likely due to the presence of topological configurations resembling that in Fig.~\ref{fig:basic:exponential}. Hence, we do not consider this algorithm in our experimental framework and focus exclusively on polynomial-time algorithms.
\subsection{Test Environment}

\begin{table*}[t]
\resizebox{\textwidth}{!}{ 
\renewcommand{\arraystretch}{0.9} % this reduces the vertical spacing between rows
\begin{tabular}{c||rr|r||rr|r||rr|r||rr|r}

\multirow{3}{*}{\textbf{graph}}& \multicolumn{3}{c||}{$k=2$}& \multicolumn{3}{c||}{$k=4$}& \multicolumn{3}{c||}{$k=8$}& \multicolumn{3}{c}{$k=16$}\\
\cline{2-13}
& \multicolumn{2}{c|}{\textsc{runtime} (s)} & \multirow{2}{*}{\shortspeedup}
& \multicolumn{2}{c|}{\textsc{runtime} (s)} & \multirow{2}{*}{\shortspeedup}
& \multicolumn{2}{c|}{\textsc{runtime} (s)} & \multirow{2}{*}{\shortspeedup}
& \multicolumn{2}{c|}{\textsc{runtime} (s)} & \multirow{2}{*}{\shortspeedup}\\
\cline{2-3}\cline{5-6}\cline{8-9}\cline{11-12}
& \ssyen & \shortpolyyen &  & \ssyen & \shortpolyyen& & \ssyen & \shortpolyyen & & \ssyen & \shortpolyyen &\\
\hline
\linuxss & 0.26 & 0.02 & 14.79 & 0.65 & 0.07 & 8.75 & 1.48 & 0.22 & 6.99 & 4.28 & 1.07 & 4.08 \\
\citatss & 2.00 & 0.16 & 12.26 & 7.31 & 0.42 & 19.11 & 26.92 & 1.38 & 19.53 & 43.53 & 12.15 & 3.61 \\
\slashdss & 3.03 & 0.22 & 14.27 & 6.21 & 0.46 & 14.23 & 19.68 & 7.64 & 2.66 & 37.88 & 14.88 & 3.70 \\
\caidass &  1046.04 & 1.25 & 931.92 & 3008.32 & 2.86 & 1264.17 & 7402.08 & 31.24 & 617.47 & 12095.90 & 117.80 & 593.08 \\
\peerss  & 24.73 & 2.30 & 10.98 & 69.97 & 18.77 & 3.75 & 175.83 & 75.74 & 2.33 & 391.66 & 229.21 & 1.71 \\
\luxss & 32510.60  & 3.27&  9566.80& 65648.57 &25.88 &3677.38 &232525.70&190.17&1222.71 &224704.82 & 1004.18 & 223.77  \\
% \wikivss & 3.78 & 1.47 & 2.59 & 10.17 & 3.17 & 3.21 & 23.67 & 6.54 & 3.62 & 57.05 & 16.94 & 3.39 \\
% \wikivss & 3.00 & 1.35 & 2.24 & 7.85 & 2.65 & 2.96 & 18.69 & 5.92 & 3.16 & 45.97 & 14.87 & 3.10 \\
\amazonss  & 138.63 & 1.91 & 72.50 & 357.48 & 3.74 & 95.53 & 639.92 & 15.92 & 38.78 & 1199.77 & 68.97 & 19.15 \\
\emaildss & 21.49 & 1.07 & 20.17 & 60.64 & 2.13 & 28.50 & 140.85 & 4.98 & 28.30 & 317.07 & 12.67 & 25.03 \\
\brightss & 50.25 & 3.17 & 15.93 & 282.77 & 6.67 & 42.45 & 369.34 & 23.32 & 15.88 & 888.25 & 83.07 & 10.80 \\ 
\youtubss & 303.58 & 3.96 & 76.88 & 1392.39 & 15.02 & 94.33 & 2067.67 & 40.85 & 53.95 & 7491.01 & 765.61 & 11.54 \\
% \barabss & 80.39 & 6.43 & 12.51 & 300.79 & 39.55 & 7.61 & 781.57 & 226.17 & 3.46 & 1724.79 & 843.02 & 2.05 \\
\barabss & 107.78 & 5.35 & 20.15 & 359.33 & 45.33 & 7.93 & 787.97 & 148.28 & 5.31 & 1536.22 & 424.09 & 3.62 \\
\mocss &  7247.56 & 2.18 & 3349.80 & 30169.19 & 5.76 & 5237.91 & 68681.53 & 17.25 & 4089.43 & 100914.23 & 113.12 & 939.80 \\
\arxivss  & 116.29 & 7.56 & 15.34 & 252.98 & 15.37 & 16.56 & 680.67 & 51.82 & 13.64 & 1222.15 & 164.00 & 8.16 \\
\ciscoss  & 1456.39 & 3.62 & 402.07 & 5240.55 & 43.25 & 121.16 & 11972.64 & 19.28 & 620.94 & 27380.17 & 406.29 & 67.39 \\

\facebss & 216.45 & 8.50 & 25.45 & 443.46 & 23.03 & 19.25 & 919.75 & 80.63 & 11.41 & 2214.98 & 316.76 & 6.99 \\

\episs & 40.25 & 4.35 & 9.20 & 94.31 & 9.68 & 9.71 & 234.33 & 21.81 & 10.69 & 1042.70 & 53.76 & 17.47 \\
% \wktss  & 29.33 & 7.67 & 3.83 & 113.80 & 16.33 & 7.00 & 276.10 & 35.63 & 7.75 & 501.21 & 80.36 & 6.23 \\
\biologss  & 920.54 & 25.10 & 36.68 & 2143.79 & 63.35 & 34.05 & 7041.90 & 313.83 & 23.79 & 16838.18 & 2500.08 & 6.80 \\
\direrdosss  & 4.76 & 3.43 & 1.39 & 13.01 & 7.11 & 1.83 & 33.00 & 18.47 & 1.83 & 83.38 & 51.90 & 1.69 \\
\Xhline{4\arrayrulewidth}
\multirow{3}{*}{\textbf{graph}}& \multicolumn{3}{c||}{$k=2$}& \multicolumn{3}{c||}{$k=4$}& \multicolumn{3}{c||}{$k=8$}& \multicolumn{3}{c}{$k=16$}\\
\cline{2-13}
& \multicolumn{2}{c|}{\textsc{runtime} (s)} & \multirow{2}{*}{\shortspeedup}
& \multicolumn{2}{c|}{\textsc{runtime} (s)} & \multirow{2}{*}{\shortspeedup}
& \multicolumn{2}{c|}{\textsc{runtime} (s)} & \multirow{2}{*}{\shortspeedup}
& \multicolumn{2}{c|}{\textsc{runtime} (s)} & \multirow{2}{*}{\shortspeedup}\\
\cline{2-3}\cline{5-6}\cline{8-9}\cline{11-12}

& \sspnc & \shortpolypnc & & \sspnc & \shortpolypnc&  & \sspnc & \shortpolypnc & & \sspnc & \shortpolypnc & \\
\hline
\linuxss & 5.22 & 0.13 & 54.73 & 5.64 & 0.57 & 10.48 & 6.09 & 1.00 & 6.43 & 7.35 & 2.04 & 3.62 \\
\citatss & 152.30 & 0.63 & 286.62 & 167.91 & 2.98 & 89.33 & 165.33 & 6.92 & 26.91 & 198.91 & 53.28 & 3.79 \\

% \slashdss  & 53.81 & 0.32 & 233.33 & 59.54 & 2.40 & 35.66 & 110.74 & 39.26 & 2.87 & 168.88 & 73.80 & 3.94 \\
\slashdss &  56.20 & 0.29 & 245.53 & 60.28 & 2.54 & 31.90 & 98.33 & 34.15 & 2.94 & 134.92 & 56.26 & 3.81 \\
\caidass &  19138.74 & 31.45 & 4918.95 & 19824.46 & 38.59 & 535.36 & 23463.01 & 146.96 & 242.71 & 28526.35 & 463.62 & 137.19 \\
\peerss  & 3135.32 & 304.80 & 10.38 & 3247.06 & 902.18 & 3.60 & 3415.10 & 1515.82 & 2.26 & 3812.58 & 2108.34 & 1.81 \\

\luxss &  25634.25 & 24.14 & 3004.36 & 28130.26 & 164.41 & 180.38 & 31067.73 & 339.07 & 99.86 & 42650.01 & 671.46 & 74.32 \\ 
% \wikivss & 895.39 & 2.01 & 446.36 & 950.73 & 5.80 & 167.90 & 922.83 & 27.26 & 34.03 & 975.71 & 79.70 & 12.45 \\
% \wikivss & 762.71 & 1.78 & 429.47 & 798.02 & 5.17 & 156.44 & 854.18 & 25.30 & 34.14 & 953.65 & 81.19 & 11.93 \\
\amazonss & 6700.12 & 5.87 & 1235.82 & 6985.21 & 10.96 & 645.14 & 8336.40 & 211.47 & 42.40 & 9061.01 & 668.95 & 18.59 \\
\emaildss & 16882.38 & 9.42 & 1999.62 & 17959.84 & 39.10 & 482.75 & 18420.20 & 101.50 & 181.75 & 21227.37 & 336.61 & 64.68 \\
\brightss & 54469.45 & 97.31 & 697.01 & 58559.41 & 117.47 & 555.36 & 62385.45 & 1487.89 & 43.38 & 72495.34 & 4527.03 & 16.13 \\

\youtubss & 15215.78 & 13.94 & 1158.48 & 16949.68 & 67.88 & 371.37 & 18787.27 & 438.53 & 50.10 & 26439.27 & 2427.20 & 12.27 \\
\barabss & 99439.21 & 3458.27 & 28.75 & 121946.11 & 14027.66 & 8.69 & 122631.33 & 28558.34 & 4.29 & 135304.94 & 55305.18 & 2.45 \\
\mocss  & 12973.91 & 2.55 & 5259.79 & 13945.51 & 6.39 & 2215.55 & 14379.65 & 73.76 & 195.35 & 15803.83 & 217.69 & 72.91 \\
\arxivss & 25407.09 & 34.45 & 1391.62 & 26276.72 & 135.41 & 217.26 & 26875.59 & 1350.72 & 29.43 & 29629.26 & 3518.56 & 10.35 \\
\ciscoss  & 19199.88 & 68.77 & 279.19 & 26491.83 & 406.70 & 65.14 & 38130.15 & 86.40 & 441.32 & 65096.57 & 1536.69 & 42.36 \\

\facebss &  29367.16 & 30.18 & 973.15 & 29120.78 & 608.44 & 47.86 & 30483.41 & 1966.75 & 15.50 & 32412.54 & 3131.96 & 10.35 \\
\episs & 37512.01 & 34.67 & 1546.97 & 43685.82 & 263.12 & 169.27 & 43349.16 & 938.64 & 49.82 & 46936.12 & 1547.90 & 30.48 \\
% \wktss  & 17762.43 & 20.59 & 876.09 & 18291.17 & 81.30 & 241.17 & 20746.55 & 356.24 & 59.35 & 21831.84 & 929.99 & 23.55 \\
\biologss  & 13237.96 & 42.40 & 342.50 & 13076.11 & 171.61 & 85.93 & 12921.53 & 599.54 & 23.25 & 13276.35 & 1220.97 & 11.54 \\
\direrdosss  & 5244.63 & 77.83 & 67.67 & 4954.62 & 319.84 & 15.50 & 5116.20 & 759.26 & 6.74 & 5202.65 & 1467.91 & 3.54 \\

\end{tabular}
}
\caption{Results of the executions of \ssyen and \shortpolyyen (top) and of \sspnc and \shortpolypnc (bottom) for $k\in\{2,4,8,16\}$.}
\label{table:results}
\end{table*}
All algorithms were implemented using NetworKit~\cite{networkit} and NetworkX~\cite{hagberg2008exploring}, two widely adopted toolkits for testing graph algorithms. Our code is written in Python, with selected performance-critical routines reimplemented in Cython and Rust. 
Experiments were carried out using Python 3.13 on a Linux system (Kernel 6.8.0-47) running on a workstation equipped with an Intel Intel Core i9-11900KF\textsuperscript{\textcopyright} clocked at 3.50GHz and 96 GB RAM.
Following previous studies on shortest-path algorithms~\cite{delling2014robust,AkibaHNIY15,DEmidioFFLP19,AngrimanGLMNPT19}, as input to our tests we employ a large dataset that includes both real-world graphs, from publicly available repositories~\cite{rossi2015network,snapnets}, and synthetic instances, generated via well-known random models (e.g., \emph{Erd\H{o}s-R\'enyi} or \textit{Barabási-Albert} models)~\cite{B01}, with heterogeneous sizes, structures, orientations, and weight distributions. Details on used inputs are given in Table~\ref{table:input}. For each dataset we extract the largest (simply or strongly) connected component.%
\subsection{Executed Tests}
For each graph of Table~\ref{table:input:subset1} and $k \in \{2, 4, 8, 16\}$ we execute \shortpolyyen, \shortpolypnc, \ssyen, and \sspnc, measuring their running times to solve \ssksispshort. Depending on whether the graph is directed or undirected, and weighted or unweighted, we use the appropriate implementation of each method to handle the specific case.
%We employ standard bidirectional traversal techniques in shortest-path computations to reduce runtime.
The choice of $k$ is inspired by real-world applications of top-$k$ simple shortest paths, where the number of required paths rarely exceeds a few tens~\cite{AkibaHNIY15,DD23,num_k_paths}.
For completeness and to evaluate the scalability of the algorithms against $k$ we also extend our experimentation to significantly larger values of $k$: specifically, for a selection of graphs (see Table~\ref{table:input:subset2}), we run and measure the running time of the four algorithms with $k \in \{2^i\}_{i=1,2,\dots,10}$. The selection of graphs is imposed by the large running times of the algorithms under study when $k$ increases substantially.
To mitigate the bias from the choice of the root, we performed three independent runs for each combination of graph and $k$, selecting three different roots uniformly at random from the graph's vertex set: we report running times averaged over these runs.
For validity, we verify the correctness of solutions to \ssksispshort calculated by our implementations by comparing them with those produced by the NetworkX standard implementation of Yen's algorithm. 

\begin{table*}[t]
\renewcommand{\arraystretch}{0.9} % this reduces the vertical spacing between rows
\resizebox{\textwidth}{!}{ 
\begin{tabular}{c||rr|r||rr|r||rr|r||rr|r||rr|r}

\multirow{3}{*}{\textbf{graph}}& \multicolumn{3}{c||}{$k=2$}& \multicolumn{3}{c||}{$k=4$}& \multicolumn{3}{c||}{$k=8$}& \multicolumn{3}{c}{$k=16$}& \multicolumn{3}{c}{$k=32$}\\
\cline{2-16}
& \multicolumn{2}{c|}{\textsc{runtime} (s)} & \multirow{2}{*}{\shortspeedup}
& \multicolumn{2}{c|}{\textsc{runtime} (s)} & \multirow{2}{*}{\shortspeedup}
& \multicolumn{2}{c|}{\textsc{runtime} (s)} & \multirow{2}{*}{\shortspeedup}
& \multicolumn{2}{c|}{\textsc{runtime} (s)} & \multirow{2}{*}{\shortspeedup}
& \multicolumn{2}{c|}{\textsc{runtime} (s)} & \multirow{2}{*}{\shortspeedup}\\
\cline{2-3}\cline{5-6}\cline{8-9}\cline{11-12}\cline{14-15}
& \ssyen & \shortpolyyen &  & \ssyen & \shortpolyyen& & \ssyen & \shortpolyyen & & \ssyen & \shortpolyyen && \ssyen & \shortpolyyen &\\
\hline
\itass  & 15.61 & 0.06 & 261.25 & 33.61 & 0.71 & 48.07 & 70.35 & 3.82 & 18.84 & 182.21 & 10.43 & 17.52 & 426.62 & 48.13 & 9.01 \\
\scfss & 0.17 & 0.02 & 11.03 & 0.48 & 0.05 & 9.61 & 1.13 & 0.20 & 6.08 & 2.89 & 0.73 & 4.13 & 7.29 & 2.35 & 3.17 \\
\bitcoinss  & 34.18 & 0.44 & 77.85 & 93.43 & 3.43 & 27.86 & 255.63 & 16.05 & 15.99 & 538.53 & 66.83 & 8.16 & 1407.71 & 264.26 & 5.33 \\
\oregonss & 12.10 & 0.33 & 36.82 & 35.36 & 0.85 & 41.82 & 129.06 & 6.21 & 34.46 & 227.04 & 14.97 & 19.00 & 593.86 & 74.54 & 8.60 \\
% \openfliss &1.01 & 0.27 & 3.88 & 2.50 & 0.47 & 5.59 & 5.64 & 1.01 & 5.74 & 15.82 & 2.59 & 6.09 & 38.38 & 7.86 & 4.86 \\
\googless & 13.06 & 0.54 & 24.23 & 37.64 & 2.94 & 12.83 & 81.29 & 14.98 & 5.58 & 189.63 & 48.66 & 3.91 & 574.69 & 177.38 & 3.22 \\
\midrule
\multirow{3}{*}{\textbf{graph}}& \multicolumn{3}{c||}{$k=64$}& \multicolumn{3}{c||}{$k=128$}& \multicolumn{3}{c||}{$k=256$}& \multicolumn{3}{c}{$k=512$}& \multicolumn{3}{c}{$k=1024$}\\
\cline{2-16}
& \multicolumn{2}{c|}{\textsc{runtime} (s)} & \multirow{2}{*}{\shortspeedup}
& \multicolumn{2}{c|}{\textsc{runtime} (s)} & \multirow{2}{*}{\shortspeedup}
& \multicolumn{2}{c|}{\textsc{runtime} (s)} & \multirow{2}{*}{\shortspeedup}
& \multicolumn{2}{c|}{\textsc{runtime} (s)} & \multirow{2}{*}{\shortspeedup}
& \multicolumn{2}{c|}{\textsc{runtime} (s)} & \multirow{2}{*}{\shortspeedup}\\
\cline{2-3}\cline{5-6}\cline{8-9}\cline{11-12}\cline{14-15}
& \ssyen & \shortpolyyen &  & \ssyen & \shortpolyyen& & \ssyen & \shortpolyyen & & \ssyen & \shortpolyyen && \ssyen & \shortpolyyen &\\
\hline
\itass & 1017.23 & 183.41 & 6.48 & 1947.49 & 532.18 & 3.97 & 3717.61 & 1388.11 & 2.83 & 11375.59 & 6425.41 & 1.79 & 28667.78 & 16576.38 & 1.74 \\
\scfss & 19.67 & 7.14 & 2.77 & 54.41 & 23.88 & 2.29 & 158.97 & 77.15 & 2.06 & 522.59 & 304.93 & 1.71 & 1898.88 & 1232.71 & 1.54 \\
\bitcoinss & 3729.68 & 1023.93 & 3.68 & 7264.18 & 2605.65 & 2.79 & 16805.06 & 6245.23 & 2.70 & 53422.27 & 20644.78 & 2.59 & 106783.87 & 42113.28 & 2.54 \\
\oregonss & 1382.01 & 268.55 & 5.28 & 3763.50 & 1065.94 & 3.66 & 6079.05 & 1885.11 & 3.29 & 18373.12 & 8258.64 & 2.27 & 41508.45 & 23215.69 & 1.79 \\
% \openfliss & 73.65 & 25.16 & 2.95 & 201.35 & 82.40 & 2.46 & 657.16 & 335.10 & 1.96 & 1528.78 & 797.86 & 1.92 & 5050.69 & 2894.65 & 1.75 \\
\googless &  973.13 & 418.53 & 2.45 & 1863.27 & 997.38 & 1.89 & 4977.50 & 3230.20 & 1.54 & 15314.60 & 10556.06 & 1.45 & 43984.99 & 30792.52 & 1.43 \\

\Xhline{4\arrayrulewidth}

\multirow{3}{*}{\textbf{graph}}& \multicolumn{3}{c||}{$k=2$}& \multicolumn{3}{c||}{$k=4$}& \multicolumn{3}{c||}{$k=8$}& \multicolumn{3}{c}{$k=16$}& \multicolumn{3}{c}{$k=32$}\\
\cline{2-16}
& \multicolumn{2}{c|}{\textsc{runtime} (s)} & \multirow{2}{*}{\shortspeedup}
& \multicolumn{2}{c|}{\textsc{runtime} (s)} & \multirow{2}{*}{\shortspeedup}
& \multicolumn{2}{c|}{\textsc{runtime} (s)} & \multirow{2}{*}{\shortspeedup}
& \multicolumn{2}{c|}{\textsc{runtime} (s)} & \multirow{2}{*}{\shortspeedup}
& \multicolumn{2}{c|}{\textsc{runtime} (s)} & \multirow{2}{*}{\shortspeedup}\\
\cline{2-3}\cline{5-6}\cline{8-9}\cline{11-12}\cline{14-15}
& \sspnc & \shortpolypnc &  & \sspnc & \shortpolypnc& & \sspnc & \shortpolypnc & & \sspnc & \shortpolypnc && \sspnc & \shortpolypnc &\\
\midrule
\itass & 18.01 & 1.07 & 28.65 & 19.36 & 2.22 & 8.82 & 26.89 & 6.47 & 4.16 & 36.04 & 10.19 & 3.54 & 72.44 & 31.13 & 2.41 \\
\scfss & 6.41 & 0.11 & 142.95 & 7.05 & 0.42 & 17.24 & 7.92 & 0.79 & 10.13 & 11.09 & 2.56 & 4.76 & 16.34 & 4.87 & 3.65 \\
\bitcoinss & 588.49 & 5.16 & 114.05 & 591.15 & 14.43 & 40.98 & 628.30 & 44.13 & 14.24 & 694.49 & 100.13 & 6.94 & 954.04 & 191.21 & 4.99 \\
\oregonss &  1081.73 & 0.37 & 3078.67 & 1199.27 & 0.83 & 1467.22 & 1358.87 & 6.36 & 478.31 & 1722.24 & 15.39 & 139.40 & 2550.65 & 76.59 & 35.29 \\ 
\googless & 2379.78 & 2.66 & 895.49 & 2465.44 & 9.40 & 262.32 & 2534.13 & 427.69 & 5.93 & 3135.56 & 718.90 & 4.36 & 3844.48 & 1221.88 & 3.15 \\
\midrule
\multirow{3}{*}{\textbf{graph}}& \multicolumn{3}{c||}{$k=64$}& \multicolumn{3}{c||}{$k=128$}& \multicolumn{3}{c||}{$k=256$}& \multicolumn{3}{c}{$k=512$}& \multicolumn{3}{c}{$k=1024$}\\
\cline{2-16}
& \multicolumn{2}{c|}{\textsc{runtime} (s)} & \multirow{2}{*}{\shortspeedup}
& \multicolumn{2}{c|}{\textsc{runtime} (s)} & \multirow{2}{*}{\shortspeedup}
& \multicolumn{2}{c|}{\textsc{runtime} (s)} & \multirow{2}{*}{\shortspeedup}
& \multicolumn{2}{c|}{\textsc{runtime} (s)} & \multirow{2}{*}{\shortspeedup}
& \multicolumn{2}{c|}{\textsc{runtime} (s)} & \multirow{2}{*}{\shortspeedup}\\
\cline{2-3}\cline{5-6}\cline{8-9}\cline{11-12}\cline{14-15}
& \sspnc & \shortpolypnc &  & \sspnc & \shortpolypnc& & \sspnc & \shortpolypnc & & \sspnc & \shortpolyyen && \sspnc & \shortpolypnc &\\
\midrule
\itass & 167.04 & 80.78 & 2.16 & 360.45 & 193.15 & 1.92 & 612.22 & 397.55 & 1.56 & 1089.68 & 769.16 & 1.42 & 2246.45 & 1832.29 & 1.23 \\
\scfss & 31.11 & 12.19 & 2.63 & 56.95 & 27.02 & 2.16 & 104.73 & 54.44 & 1.94 & 230.52 & 142.83 & 1.62 & 474.47 & 332.20 & 1.43 \\
\bitcoinss  & 1374.11 & 460.84 & 2.98 & 3873.14 & 1601.16 & 2.42 & 6353.60 & 2786.40 & 2.28 & 8467.54 & 4432.84 & 1.91 & 18418.47 & 10017.47 & 1.84 \\\oregonss & 3948.19 & 270.99 & 15.56 & 8475.54 & 1082.66 & 8.14 & 13052.88 & 1948.15 & 6.79 & 36193.34 & 8397.07 & 4.31 & 76662.79 & 22372.69 & 3.40\\ 
\googless & 5400.40 & 1853.90 & 2.91 & 7239.20 & 4101.81 & 1.76 & 15800.12 & 10926.51 & 1.45 & 31077.07 & 20510.12 & 1.52 & 66743.96 & 44410.96 & 1.50

\end{tabular}
}
\caption{Results of the executions of \ssyen and \shortpolyyen (top) and of \sspnc and \shortpolypnc (bottom) for $k \in \{2^i\}_{i=1,2,\dots,10}$.}
\label{table:k:scale}
\end{table*}
\subsection{Analysis}%
The results of our experiments for $k\in \{2,4,8,16\}$ are summarized in Table~\ref{table:results}. For each graph and $k$, we report the average running time (in seconds) spent by \ssyen (\sspnc, respectively) and \shortpolyyen (\shortpolypnc, respectively) to solve an instance of \ssksispshort, along with the average \textit{speed-up} (column \shortspeedup), defined as the average ratio of the runtime of \ssyen (\sspnc, respectively) to the runtime of \shortpolyyen (\shortpolypnc, respectively). This ratio quantifies how much the latter is faster than the former at solving \ssksispshort. 
The main conclusion that can be drawn from the data of Table~\ref{table:results} is that both versions of Algorithm \polysinglesource  outperform the two baselines \ssyen and \sspnc in handling \ssksispshort. 
In particular, \shortpolyyen is always faster than \ssyen in all tested instances. Similarly, \shortpolypnc outperforms \sspnc.
In more detail, the runtime of \ssyen is observed to be larger than that of \shortpolyyen by factors that range from a minimum of $1.39$ (instance \direrdosss, $k=2$) to a maximum of several thousands (e.g, graphs \luxss for $k=2$ or \caidass for $k=4$ or \mocss for $k=8$).
Similarly, the measured runtime of \sspnc is always larger than that of \shortpolypnc, by factors that range from a minimum of $3.15$ (instance \linuxss, $k=16$) to a maximum of thousands (e.g. instance \youtubss for $k=2$).
In general, in the majority of our experiments Algorithms \shortpolyyen and \shortpolypnc are more than one order of magnitude faster than the respective baselines, i.e. \ssyen and \sspnc. 
Even more remarkably, the execution of the two variants of \polysinglesource often terminates within a few seconds and never lasts more than around an hour and a half, while \ssyen or \sspnc can take several hours to solve a single instance of \ssksispshort even for small $k$ (see, e.g., graph \luxss for $k=2$ or \ciscoss with $k=8$). This suggests that \polysinglesource scales better w.r.t. graph size and can be successfully used in real-world applications that manage large graphs (for $k\ll n$), while \ssyen or \sspnc are impractical even for medium-sized graphs and small $k$.
This effectiveness of \polysinglesource is also supported by the data in Table~\ref{table:k:scale} where we report the runtime of all algorithms when applied on smaller graphs but with substantially larger values of $k$, and the speed-up provided by our new solution. This experiment was designed to evaluate how the speed-up varies with $k$. Values of such parameter here were selected by a doubling strategy to highlight observed trends~\cite{McGeoch}.
From such data, we notice that the speed-up tends to decrease as $k$ increases; this is fairly expected since, as $k$ grows, the number of general predecessors that are not saturated during the execution of \polysinglesource (see line~\ref{line:notssatcalc}) increases, and, with it, the number of times a solution to \spksispshort has to be computed increases, tending to the upper bound of $n-1$. This is clearly related to the fact that, topologically speaking, the number of prefixes, shared by top-$k$ simple shortest paths from the root to different vertices, tends to decrease as $k$ increases. Hence, computing solutions together, as done by algorithms \shortpolyyen and \shortpolypnc, tends to become less effective with $k$ increasing.
Still, \shortpolyyen and \shortpolypnc remain faster than \ssyen and \sspnc even for the largest graphs and the highest values of $k$. 
Other empirical insights can be derived by comparing the algorithms' runtimes for a fixed graph and value of $k$.
For example, for $k\in\{2,4,8,16\}$ (see Table~\ref{table:results}), in weighted digraphs (e.g. \luxss or \mocss) \sspnc runs faster than \ssyen while \shortpolypnc is slightly slower than \shortpolyyen, hence a slightly reduced speed-up is observed.
Viceversa, in all other cases, \ssyen and \shortpolyyen exhibit smaller runtimes compared to \sspnc and \shortpolypnc.
The observed performance changes significantly for larger $k$ (see Table~\ref{table:k:scale}), where \pnc benefits of the precomputed shortest-path tree to reduce the number of shortest-path computations necessary to complete a solution to \spksispshort. This results in shorter runtimes for both \sspnc and \shortpolypnc when $k\geq 64$, suggests that these algorithms should be used for large $k$, and confirms that \pnc is better than Yen's for \spksispshort when $k$ increases.
\footnote{All our source codes and used inputs can be accessed at \url{https://tinyurl.com/3uew6rx9}.}

\section{Conclusion}
\label{sec:conclusion}
In this paper, we studied the \ssksisp problem (\ssksispshort) and introduced \polysinglesource, the first algorithm specifically designed to solve it, based on a careful analysis of some key structural properties of the problem's solutions.
We proved that the time complexity of \polysinglesource matches that of the best-known existing methods. However, extensive experiments show that it significantly outperforms such methods in practice, often by orders of magnitude, making it the preferred choice to handle \ssksispshort in real-world scenarios.

A promising direction for future work is to improve \polysinglesource either asymptotically, by reducing its time complexity to below $\bigO(n)$ times that of the best \spksispshort algorithm, or empirically, by enhancing its practical performance through heuristics. To support this, we plan to extend our experimental analysis to better identify the algorithm's computational bottlenecks.
Finally, a major open question is whether the strategy underlying \polysinglesource can be generalized to efficiently compute the $k$ simple shortest paths for all vertex pairs. Currently, no polynomial-time algorithm is known, for general values of $k$, which needs $o(\binom{n}{2})$ individual \spksispshort computations. Addressing this challenge would mark a significant advancement in the field.

\section*{Acknowledgments}
This work was supported by: National Group for Scientific Computation of Istituto Nazionale di Alta Matematica (GNCS-INdAM); and European Union under the Italian National Recovery and Resilience Plan of NextGenerationEU, partnership on "Telecommunications of the Future" (program RESTART), project MoVeOver/SCHEDULE ("Smart interseCtions witH connEcteD and aUtonomous vehicLEs", CUP J33C22002880001).

\clearpage
\bibliographystyle{plain}
\bibliography{biblio_cut}

\end{document}